# Second-order discontinuous Galerkin flood model: comparison with industry-standard finite volume models

Janice Lynn Ayog, Georges Kesserwani*, James Shaw, Mohammad Kazem Sharifian, Domenico Bau

*Department of Civil and Structural Engineering, University of Sheffield, Mappin St, Sheffield City Centre, Sheffield S1 3JD, UK*

**Abstract**

Finite volume (FV) numerical solvers of the two-dimensional shallow water equations are core to industry-standard flood models. The second-order Discontinuous Galerkin (DG2) alternative, although a viable way forward to improve current FV-based flood models, is yet under-studied and rarely used to support flood modelling applications. This paper systematically explores and compares the predictive properties of a robust DG2 flood model to those of prominent FV-based industrial flood models. To identify the simplest and most efficient DG2 configuration suitable for flood inundation modelling, two variants – with and without local slope limiting – are considered. The numerical conservation properties of the DG2 variants are compared to those of a first-order FV (FV1) and a second-order FV (FV2) counterparts. The DG2 variants are then tested over five realistic flooding scenarios, recommended by the UK Environment Agency to validate 2D flood model capabilities, while comparing their performance against that of four FV-based commercial models (i.e. TUFLOW-FV1, TUFLOW-FV2, TUFLOW-HPC and Infoworks ICM). Results reveal that the DG2 variant without local limiting (DG2-NL) is capable to simulate shockless flood flows featured in a wide range of flood modelling applications. The DG2-NL shows closer predictions to commercial model outputs at twice-coarser spatial resolution, and can run twice faster to produce more informative hydrograph with small-scale transients over long-range simulations, even when the sampling is far away from the flooding source.

*Corresponding author.
E-mail address: g.kesserwani@shef.ac.uk (G. Kesserwani)

# 1. Introduction

In the past decades, the finite volume (FV) method has emerged as a core basis to the numerical solvers of the full two-dimensional (2D) shallow water equations (SWE). These solvers have been adopted and enhanced in the current industrial flood modelling packages (Alcrudo, 2004; Neelz and Pender, 2009; Teng et al., 2017) and are often preferred by government agencies in the UK, Netherlands and Australia for flood risk assessment and for the management of urban and rural floodplains (Engineers Australia, 2012; Henckens and Engel, 2017; Neelz and Pender, 2013). FV-based flood models are known for their ability to capture the widest range of flow transitions, making them well-suited to provide reliable predictions for complex, real-world flood applications. For example, FV models are applied to produce hydrographs with detailed velocity transients to estimate structural damages on residential buildings (Pistrika and Jonkman, 2010), to identify safe parking locations for emergency vehicles during flood evacuation (Arrighi et al., 2019), and to develop quantifiable hazard classification for flood vulnerability assessment (Costabile et al., 2020; Shirvani et al., 2020a, 2020b). Similarly, for gradually propagating floods, hydrographs produced by the FV-based models are used to estimate arrival times and flood levels in low-lying areas (Alkema, 2007; Latrubesse et al., 2020), e.g. to assess the clearance time for flood evacuation (Cheng et al., 2011) and to identify zones for flood rescue prioritisation (Patel et al., 2017).

One issue with industry-standard FV models is that they often employ a first-order accurate finite volume (FV1) solver, as in the case, for example, of TUFLOW-FV1 (BMT-WBM, 2016), Infoworks ICM (Lhomme et al., 2010), JFlow+ (Crossley et al., 2010), LISFLOOD-FP with Roe solver (Neal et al., 2012), RiverFlow2D (Hydronia LLC, 2019), and BASEMENT (Vetsch et al., 2018). Such models may not be ideal for certain applications, as the outputs of an FV1 solver can be severely affected by rapid accumulation of numerical diffusion, particularly when the fine resolution needed to alleviate these errors is unaffordable (Lhomme et al., 2010; Neal et al., 2012). For example, an FV1-based flood model tends to predict late arrival times and narrower wetting extents (Schubert et al., 2008; Kesserwani and Sharifian 2020), and fails to capture small-scale transients in hydrograph predictions (Soares-Frazão and Zech, 2008; de Almeida et al., 2018). Error analyses in the published literature show that FV1-based models tend to produce larger deviations in hydrograph predictions when gauging stations are located far from the inflow and the flood duration lasts several days (Echeverribar et al., 2019; Horváth et al., 2020; Xia et al., 2019).

To improve the capability of FV-based models, second-order accurate finite volume (FV2) solvers have been used in existing industrial flood models, for example in TUFLOW-FV2 (BMT-WBM, 2016), TUFLOW-HPC (BMT-WBM, 2018), ANUGA (Mungkasi and Roberts, 2013), and Iber (Bladé et al., 2014). Many FV2-based models adopt the Monotonic Upstream-centred Scheme for Conservation Laws (MUSCL) approach to reconstruct piecewise-linear solutions, but this widens the calculation stencil to the neighbour's neighbour of the element where local flow data are updated. The MUSCL approach comes hand-in-hand with a Total Variation Diminishing (TVD) slope limiter to ensure that the reconstructed flow solutions are free from unphysical oscillations. As the slope limiter is inherent to any MUSCL-FV2 solver and is applied globally over each grid element, a key challenge has been to choose a slope limiter function that can simultaneously retain second-order accurate, oscillatory-free solutions, and ensure reliable reproduction of real-world features (Bai et al., 2018; Guard et al., 2013; Reis et al., 2019; Sanders and Bradford, 2006; Yoon and Kang, 2004). This level of fidelity is particularly important in flood inundation modelling given the presence of moving wet-dry fronts where MUSCL-FV2 outputs could be significantly affected, such as in the cases of poor capturing of vanishing velocities (Delis et al., 2011; Delis and Nikolos, 2013; Hou et al., 2015; Zhao et al., 2018; Bai et al, 2018) and wave arrival times if the resolution is insufficiently fine (Kesserwani and Wang, 2014; Zhao et al., 2018).

The second-order discontinuous Galerkin (DG2) method provides an alternative approach to generate and update a piecewise-planar solution per grid element, where the solution slopes are defined intrinsically, thereby avoiding slope reconstruction and reducing the calculation stencil to the direct neighbours. With a DG2 solver, slope limiting becomes irrelevant and can drastically spoil the quality the predictions if applied globally as with MUSCL-FV2. Rather, the slope limiter should only be applied locally to stabilise the DG2 solution at very steep discontinuities within the wet portions of the computational domain (Kesserwani and Liang, 2012a; Krivodonova et al., 2004). Hence, the choice for the limiter function is not pertinent with a DG2 solver, for which the key challenge has rather been to identify and apply a method to localise the operation of the slope limiter, which increases the computational cost of a DG2 solver (Fu and Shu, 2017; Le et al., 2020; Marras et al., 2018; Qiu and Shu, 2005; Vater et al., 2019). To date, very few 2D hydraulic modelling packages have adopted DG-based solvers, of which DG-SWEM (Bunya et al., 2009; Kubatko et al., 2006), SLIM (Lambrechts et al., 2010) and Thetis (Kärnä et al., 2018) are among the examples. Existing DG-based solvers have been primarily aimed to support modelling applications in lake, estuary and coastal systems (Le Bars et al., 2016; Clare et al., 2020; Le et al., 2020; Mulamba et al., 2019; Pham Van et al., 2016; Wood et al., 2020). However, the application of DG2 for 2D flood inundation modelling has been little explored (Kesserwani and Sharifian, 2020; Kesserwani and Wang, 2014), and a dedicated study is needed to identify: (i) the optimal DG2 configuration for flood modelling applications, and (ii) the potential capabilities of DG2 compared to commonly used FV-based alternatives, including FV-based industry-standard models.

This paper presents such a study for a robust grid-based DG2 solver designed for flood modelling applications, which considers two possible variants: with and without local slope limiting. The variants of the DG2 solver are diagnostically assessed – alongside a standard MUSCL-FV2 solver and its FV1 variant – to identify in particular their potentials in modelling flow features relevant to flood inundation, as well as their mass and energy conservation properties over time. The DG2 variants are then benchmarked against the outputs of four industry-standard FV-based flood models at similar resolution, and against the outputs of the best performer DG2 variant and the MUSCL-FV2 solver at twice-coarser resolution. The comparisons include qualitative and quantitative analyses aimed at identifying when a DG2-based flood model could add value to support industrial-scale flood modelling, and which DG2 variant is the most appropriate in relation to the intended modelling study.

Accompanying model data are available on Zenodo (Ayog and Kesserwani, 2020), and the details to freely access the DG2 solver software are provided in the acknowledgements section.

**2. Grid-based flood models**

The slope-decoupled DG2 solver has been designed to provide well-balanced piecewise-planar solutions on a simplified stencil that is compatible with that of existing grid-based FV flood models. It has been demonstrated to reduce the runtime cost per element by 2.6 times compared to a standard DG2 solver while preserving second-order accuracy (Kesserwani et al., 2018). In this section, this DG2 solver (Sec. 2.1) is optimised further to make it suitable for 2D flood inundation modelling. A MUSCL-FV2 alternative (Sec. 2.2) is presented and its main differences and similarities to the DG2 solver are discussed. The conservation properties of both DG2 and MUSCL-FV2 solvers are then analysed (Sec. 2.3) through a synthetic test case involving nonlinear flow with periodic wetting and drying cycles over an uneven topography.

The DG2 and MUSCL-FV2 solvers rely on the Godunov-type approach to numerically solve the two-dimensional (2D) shallow water equations (SWE), expressed in a conservative form:

$$\partial_t \mathbf{U} + \partial_x \mathbf{F}(\mathbf{U}) + \partial_y \mathbf{G}(\mathbf{U}) = \mathbf{S}(\mathbf{U}) \qquad (1)$$

where $\mathbf{U}(x, y, t) = [h, hu, hv]^T$ is the flow vector at location $(x, y)$ and time $t$, including the water depth $h$ (m) and the components of the unit-width flow discharge $hu$ (m²/s) and $hv$ (m²/s), with $u$ (m/s) and $v$ (m/s) being the components of the depth-averaged horizontal velocity. The physical flux has the components $\mathbf{F} = [hu, (hu)^2 h^{-1} + 0.5gh^2, huv]^T$ and $\mathbf{G} = [hv, huv, (hv)^2 h^{-1} + 0.5gh^2]^T$ in which $g$ (m/s²) is the gravity acceleration. The source term vector $\mathbf{S}$ is decomposed into $\mathbf{S} = \mathbf{S}_b + \mathbf{S}_f$, with $\mathbf{S}_b = [0, -gh\partial_x z, -gh\partial_y z]^T$ including the bed-slope terms of a given 2D topography $z$, and $\mathbf{S}_f = [0, -C_f u\sqrt{u^2 + v^2}, -C_f v\sqrt{u^2 + v^2}]^T$ including the friction effects. $C_f = g\, n_M^2\, h^{-1/3}$ is a drag term expressed in terms of the Manning's roughness coefficient $n_M$ (s⁻¹ m^{1/3}).

Both numerical solvers are aimed to solve Eq. (1) on a 2D grid made of $M \times N$ uniform rectangular elements $Q_c$ ($c = 1, …, M \times N$). Each grid element $Q_c$, centred at $(x_c, y_c)$ and with size $\Delta x \times \Delta y$, can be expressed as $Q_c = [x_c - \frac{\Delta x}{2}, x_c + \frac{\Delta x}{2}] \times [y_c - \frac{\Delta y}{2}, y_c + \frac{\Delta y}{2}]$ (see Fig. 1).

## 2.1. DG2 solver

### 2.1.1. Local piecewise-planar solutions

The DG2 method uses local piecewise-planar approximate solutions, $\mathbf{U_h}$, to the flow vector $\mathbf{U}$. Over each grid element $Q_c$, $\mathbf{U_h}$ is expanded by coefficients representing an average, $\mathbf{U}_c^0(t)$, and two directionally independent slopes, $\mathbf{U}_c^{1x}(t)$ and $\mathbf{U}_c^{1y}(t)$, as follows:

$$\mathbf{U_h}(x,y,t)|_{Q_c} = \mathbf{U}_c^0(t) + \frac{(x-x_c)}{\Delta x/2}\mathbf{U}_c^{1x}(t) + \frac{(y-y_c)}{\Delta y/2}\mathbf{U}_c^{1y}(t) \qquad (2)$$

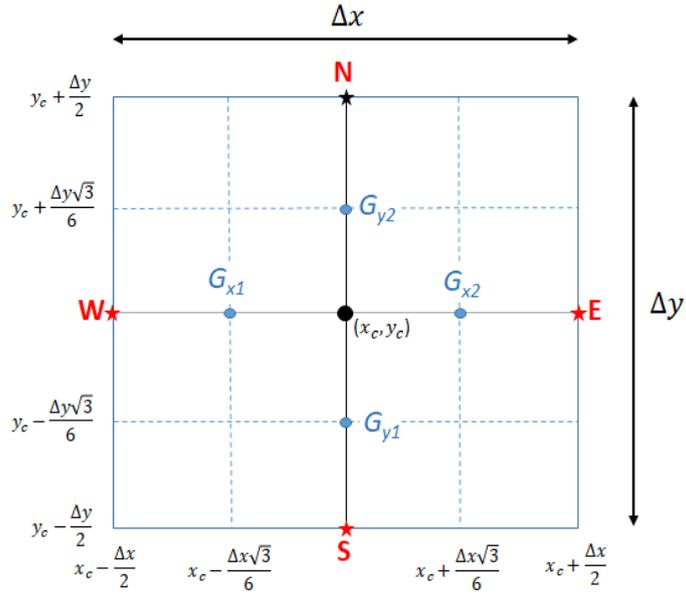

**Fig. 1.** A rectangular element $Q_c$ on which the slope-decoupled DG2 form (Sec. 2.1) is applied. Points $E$, $W$, $N$ and $S$ denote the face-centred nodes at the eastern, western, northern and southern faces. Points $G_{xi}$ and $G_{yi}$ ($i = 1, 2$) denote Gaussian evaluation points along the $x$- and $y$-directional centreline, respectively.

$\mathbf{U}_c^0(t)$ $\mathbf{U}_c^{1x}(t)$ and $\mathbf{U}_c^{1y}(t)$ are also referred to as DG2 modes, which are initialised from a given initial condition, $\mathbf{U_0}(x,y,t)$, such that:

$$\mathbf{U}_c^0(0) = \tfrac{1}{2}[\mathbf{U_0}(E) + \mathbf{U_0}(W)] = \tfrac{1}{2}[\mathbf{U_0}(N) + \mathbf{U_0}(S)] \qquad (3a)$$

$$\mathbf{U}_c^{1x}(0) = \tfrac{1}{2}[\mathbf{U_0}(E) - \mathbf{U_0}(W)] \qquad (3b)$$

$$\mathbf{U}_c^{1y}(0) = \tfrac{1}{2}[\mathbf{U_0}(N) - \mathbf{U_0}(S)] \qquad (3c)$$

where $\mathbf{U_0}(p)$, $p = E, W, N$ and $S$, are the initial condition values at the face-centred nodes (Fig. 1), which are preliminarily averaged from the corner node values (see Eqs. 72-81 in Kesserwani et al., 2018). The time-invariant modes, $z_c^0$, $z_c^{1x}$ and $z_c^{1y}$, which represent the piecewise-planar terrain approximations, $z_h(x,y)$, are initialised in a fashion similar to Eqs. (3) with the values at the corner nodes representing the data points extracted from a Digital Elevation Model (DEM).

**2.1.2. Well-balanced and depth-positivity preserving DG2 operators**

In an explicit second-order Runge-Kutta (RK2) time integration scheme, with a maximum Courant number $Cr = 0.3$ (Cockburn and Shu, 2001), the DG2 modes of the flow vector, $\mathbf{U}_c^K(t)$, are updated by solving three independent ordinary differential equations (ODEs), $\partial_t \mathbf{U}_c^K(t) = \mathbf{L}_c^K$, with $\mathbf{L}_c^K$ being spatial operators ($K = 0, 1x, 1y$). These spatial DG2 operators (see Eqs. 40-42 in Kesserwani et al. (2018)) – after algebraic manipulations and zeroing cross-dimensional bed-slope dependencies – can be reduced to read:

$$\mathbf{L}_c^0 = -\frac{1}{\Delta x}\left(\widetilde{\mathbf{F}}_E - \widetilde{\mathbf{F}}_W\right) - \frac{1}{\Delta y}\left(\widetilde{\mathbf{G}}_N - \widetilde{\mathbf{G}}_S\right) - 2g \begin{bmatrix} 0 \\ \frac{\bar{h}_c^{0x} \bar{z}_c^{1x}}{\Delta x} \\ \frac{\bar{h}_c^{0y} \bar{z}_c^{1y}}{\Delta y} \end{bmatrix} \tag{4a}$$

$$\mathbf{L}_c^{1x} = -\frac{3}{\Delta x}\left\{\left(\widetilde{\mathbf{F}}_E + \widetilde{\mathbf{F}}_W\right) - \left(\mathbf{F}\left(\bar{\mathbf{U}}_c^{0x} + \frac{1}{\sqrt{3}}\bar{\mathbf{U}}_c^{1x}\right) + \mathbf{F}\left(\bar{\mathbf{U}}_c^{0x} - \frac{1}{\sqrt{3}}\bar{\mathbf{U}}_c^{1x}\right)\right) + \frac{2g}{3}\begin{bmatrix} 0 \\ \bar{h}_c^{1x} \bar{z}_c^{1x} \\ 0 \end{bmatrix}\right\} \tag{4b}$$

$$\mathbf{L}_c^{1y} = -\frac{3}{\Delta y}\left\{\left(\widetilde{\mathbf{G}}_N + \widetilde{\mathbf{G}}_S\right) - \left(\mathbf{G}\left(\bar{\mathbf{U}}_c^{0y} + \frac{1}{\sqrt{3}}\bar{\mathbf{U}}_c^{1y}\right) + \mathbf{G}\left(\bar{\mathbf{U}}_c^{0y} - \frac{1}{\sqrt{3}}\bar{\mathbf{U}}_c^{1y}\right)\right) + \frac{2g}{3}\begin{bmatrix} 0 \\ 0 \\ \bar{h}_c^{1y} \bar{z}_c^{1y} \end{bmatrix}\right\} \tag{4c}$$

Note that the friction effects are excluded from the spatial operators in Eqs. (4) as they are added separately (Sec. 2.1.3). The terms $\widetilde{\mathbf{F}}_E$, $\widetilde{\mathbf{F}}_W$, $\widetilde{\mathbf{G}}_N$ and $\widetilde{\mathbf{G}}_S$ are the inter-elemental fluxes evaluated from the limits of the piecewise-planar approximate solutions at either side of the face-centred nodes $p = E, W, N$ and $S$ (Fig. 1) after applying appropriate revisions to preserve well-balancedness and depth-positivity (Liang and Marche, 2009). For example, the limits at the eastern face-centred node, $E$, are expressed as: $\mathbf{U}_E^- = \mathbf{U}_c^0(t) + \mathbf{U}_c^{1x}(t)$ and $\mathbf{U}_E^+ = \mathbf{U}_{nei_E}^0(t) - \mathbf{U}_{nei_E}^{1x}(t)$, where '$nei_E$' is the index of the direct neighbour from the eastern side. Both limits at $E$ for $z_h$ are equal, $z_E = z_h(E^\pm)$, as Eq. (3) ensures the continuity property (Kesserwani et al., 2018). Dummy limits that are depth-positivity preserving, denoted with the '*' symbol, are reconstructed for the components of the flow vector, as: $h_E^{\pm,*} = \max(0, h_E^\pm)$, $(hu)_E^{\pm,*} = h_E^{\pm,*} u_E^\pm$ and $(hv)_E^{\pm,*} = h_E^{\pm,*} v_E^\pm$, where $u_E^\pm = (hu)_E^\pm / h_E^\pm$ and $v_E^\pm =$ $(hv)_E^\pm / h_E^\pm$ when $h_E^\pm$ is above a drying tolerance, $tol_{dry}$, and zeroed when $h_E^\pm \leq tol_{dry}$. A dummy revision of the continuous topography limit $z_E$ is also applied to preserve well-balancedness when a motionless body of water is blocked by a high wall, i.e. $z_E^* = z_E - \max(0, -h_E^-)$. The revised limits $\mathbf{U}_E^{\pm,*}$ are then used to evaluate $\tilde{\mathbf{F}}_E = \tilde{\mathbf{F}}(\mathbf{U}_E^{-,*}, \mathbf{U}_E^{+,*})$ via a two-argument numerical flux function, $\tilde{\mathbf{F}}$, based on an approximate Riemann solver. Here, the Riemann solver of Roe is applied as it forms the basis of the flood models explored in Sec. 3 and it is arguably the best choice with a DG2 solver for the SWE with source terms (Kesserwani et al., 2008). The fluxes $\tilde{\mathbf{F}}_W$, $\tilde{\mathbf{G}}_N$ and $\tilde{\mathbf{G}}_S$ are evaluated by analogy, after reconstructing dummy flow limits ($\mathbf{U}_W^{\pm,*}$, $\mathbf{U}_N^{\pm,*}$ and $\mathbf{U}_S^{\pm,*}$) and topography limits ($z_W^*$, $z_N^*$ and $z_S^*$). Dummy DG2 modes are then formed, appended with a 'bar', by reapplying Eq. (3) to the dummy nodal limits of the flow and topography, as:

$$\overline{\mathbf{U}}_c^{0x} = \tfrac{1}{2}\left(\mathbf{U}_E^{-,*} + \mathbf{U}_W^{+,*}\right) \tag{5a}$$

$$\overline{\mathbf{U}}_c^{0x} = \tfrac{1}{2}\left(\mathbf{U}_N^{-,*} + \mathbf{U}_S^{+,*}\right) \tag{5b}$$

$$\overline{\mathbf{U}}_c^{1x} = \tfrac{1}{2}\left(\mathbf{U}_E^{-,*} - \mathbf{U}_W^{+,*}\right) \text{ and } \bar{z}_c^{1x} = \tfrac{1}{2}(z_E^* - z_W^*) \tag{5c}$$

$$\overline{\mathbf{U}}_c^{1y} = \tfrac{1}{2}\left(\mathbf{U}_N^{-,*} - \mathbf{U}_S^{+,*}\right) \text{ and } \bar{z}_c^{1y} = \tfrac{1}{2}(z_N^* - z_S^*) \tag{5d}$$

Replacing the evaluated fluxes $\tilde{\mathbf{F}}_E$, $\tilde{\mathbf{F}}_W$, $\tilde{\mathbf{G}}_N$ and $\tilde{\mathbf{G}}_S$ alongside the positivity-preserving modes in Eq. (5), altogether in Eq. (4), lead to well-balanced evaluations for the DG2 operators without spurious momentum errors across wet-dry fronts located at very steep bed-slopes. Theoretical and diagnostic verification of these aspects can be found in Kesserwani et al. (2018).

### 2.1.3. Friction source term and local limiting treatments

Piecewise-planar approximate friction term integration is applied, involving the average, $(\mathbf{S}_f)_c^0$, and the two slope coefficients, $(\mathbf{S}_f)_c^{1x}$ and $(\mathbf{S}_f)_c^{1y}$. The contribution of the friction term coefficients is added to their respective discharge coefficients prior to each RK2 step, based on a standard implicit splitting integration (Kesserwani and Liang, 2010). The integration procedure is performed at the element centre $(x_c, y_c)$ to increment the discharges in $\mathbf{U}_c^0$, whereas evaluations at Gaussian points $G_{xi}$ and $G_{yi}$ (Fig. 1), followed by appropriate differentiations, are made to increment the discharges in $\mathbf{U}_c^{1x}$ and $\mathbf{U}_c^{1y}$.

Local limiting (LL) is applied as a separate step before each RK2 time stage, but after applying friction effects and boundary conditions. Slope limiting to coefficients $\mathbf{U}_c^{1x}(t)$ and $\mathbf{U}_c^{1y}(t)$ over $Q_c$ comes into play as an FV tool in order to limit the variations of these slopes with respect to the slopes differentiated of the average coefficients, based on a Total Variation Diminishing (TVD) slope limiter (Sweby 1983; Toro 1999). It is well-known that TVD slope limiting with DG methods is only useful at the elements where the solution is about to develop shock-like gradients, otherwise, it can greatly destroy the solution or even affect overall robustness (see examples in Kesserwani and Liang (2012, 2011)). With DG2 numerical solvers, LL is commonly used for which the main challenge has been to deploy a localisation procedure to identify where slope limiting is needed (Fu and Shu, 2017). The *shock detector* proposed by Krivodonova et al. (2004) is selected, as it is recognised to be one of the best options available to achieve the localisation procedure (Qiu and Shu, 2005). The shock detector is applied component-wise, per slope coefficient and physical component. For example, to detect if the $x$-directional slope component $\mathbf{U}_c^{1x}(t)$ at an element $Q_c$ needs limiting, the normalised magnitude of solution discontinuities at the face-centred eastern and western nodes, $\mathbf{DS}_p$ ($p = E, W$), need first to be calculated as:

$$\mathbf{DS}_p = \frac{|\mathbf{U}_p^+ - \mathbf{U}_p^-|}{\frac{\Delta x}{2}\max\left(\left|\mathbf{U}_c^0(t) - \frac{1}{\sqrt{3}}\mathbf{U}_c^{1x}(t)\right|, \left|\mathbf{U}_c^0(t) + \frac{1}{\sqrt{3}}\mathbf{U}_c^{1x}(t)\right|\right)} \tag{6}$$

Then, slope limiting will only be needed for $\mathbf{U}_c^{1x}(t)$ when at least one inter-elemental discontinuity is detected, namely when $\max(\mathbf{DS}_E, \mathbf{DS}_W) > 10$ (Krivodonova et al., 2004), to ensure that Eq. (6) will only detect strong discontinuities, i.e. of shock types. If LL is needed, the *Generalised minmod limiter* (Cockburn and Shu 2001) is then applied to limit the slope coefficient $\widehat{\mathbf{U}}_c^{1x}(t)$:

$$\widehat{\mathbf{U}}_c^{1x}(t) = minmod\left(\mathbf{U}_c^{1x}(t), \mathbf{U}_{nei_E}^0(t) - \mathbf{U}_c^0(t), \mathbf{U}_c^0(t) - \mathbf{U}_{nei_W}^0(t)\right) \tag{7}$$

Note that the limiter in Eq. (7) is commonly used with DG methods for LL, and that the choice of the slope limiter function is insignificant here as the limiting operation is restricted to the locality of a shock. Worth also noting that shock detection and limiting for the water depth slope component, $h_c^{1x}(t)$, is avoided, but Eqs. (6-7) is applied instead for the free-surface elevation $(h_c^{1x}(t) + z_c^{1x})$ to get $\widehat{(h_c^{1x}(t) + z_c^{1x})}$ and then deduce $\hat{h}_c^{1x}(t) = \widehat{(h_c^{1x}(t) + z_c^{1x})} - z_c^{1x}$. This ensures that the presence of terrain steps does not falsely trigger slope limiting. In a similar way, shock detection and limiting are applied to analyse the $y$-directional slope component $\mathbf{U}_c^{1y}(t)$ at $Q_c$ and potentially limit it to become $\hat{\mathbf{U}}_c^{1y}(t)$.

Regardless of slope limiting, the localisation procedure entails an overhead cost, since Eq. (6) needs to be applied twice for each element $Q_c$, and since the presence of at least one wet-dry front needs to be checked for deactivating the limiter. Otherwise Eq. (7) may unnecessarily replace $\mathbf{U}_c^{1x}(t)$ by an unphysically amplified slope that can trigger instabilities. In this version of the DG2 model, dry elements are flagged only by analysing the average coefficient of the water depth component, $h_c^0$, relative to a $tol_{dry} = 10^{-5}$ instead of analysing all local evaluations at the four Gaussian points (see Fig. 1) spanning the full piecewise-planar solution (Kesserwani et al., 2018). This measure for dry element detection is consistent with the way dry elements are detected in FV-based approaches.

**2.2. MUSCL-FV2 solver**

MUSCL-FV2 and FV1 model counterparts are described for the same computational stencil used to describe the slope-decoupled DG2 flood model (Fig. 1), while adopting similar notations to map the fundamental similarities and differences among the DG2 and FV-based flood models.

**2.2.1. Local piecewise-constant solutions**

The FV-based flood models only involve piecewise-constant averaged data, $\mathbf{U}_c^0(t)$ and $z_c^0$, per grid element $Q_c$, namely $\mathbf{U_h}(x, y, t)|_{Q_c} = \mathbf{U}_c^0(t)$ and $z_h(x, y)|_{Q_c} = z_c^0$, which are similarly initialised by Eq. (3a). Here, only one ODE needs solving, $\partial_t \mathbf{U}_c^0 = \mathbf{L}_c^0$, which is carried out with an RK2 scheme with MUSCL-FV2, and with an forward-Euler scheme with FV1, with $\mathbf{L}_c^0$ expressed as:

$$\mathbf{L}_c^0 = -\frac{1}{\Delta x}\big(\widetilde{\mathbf{F}}_E - \widetilde{\mathbf{F}}_W\big) - \frac{1}{\Delta y}\big(\widetilde{\mathbf{G}}_N - \widetilde{\mathbf{G}}_S\big) - g\begin{bmatrix}0\\ \bar{h}_c^{0x}\bar{z}_c^{1x}\\ \bar{h}_c^{0y}\bar{z}_c^{1y}\end{bmatrix} \qquad (8)$$

The fluxes $\widetilde{\mathbf{F}}_E$, $\widetilde{\mathbf{F}}_W$, $\widetilde{\mathbf{G}}_N$ and $\widetilde{\mathbf{G}}_S$ are evaluated after applying the MUSCL approach (van Leer, 1979) to form the solution limits at either side of the face-centred nodes $p$ ($p = E, W, N$ and $S$), with a different approach to produce revised limits in order to accommodate the discontinuous nature of $z_b$ in the FV-based flood models (Sec. 2.2.2).

**2.2.2. Well-balancedness and depth-positivity preserving operations**

Applying the MUSCL linear reconstructions, solution limits can be evaluated at the eastern face-centred node $E$ as:

$$\mathbf{U}_E^- = \mathbf{U}_c^0(t) + 0.5\boldsymbol{\nabla}_c^{1x} \qquad \text{and} \qquad \mathbf{U}_E^+ = \mathbf{U}_{nei_E}^0(t) - 0.5\boldsymbol{\nabla}_{nei_E}^{1x} \qquad (9)$$

where $\boldsymbol{\nabla}_c^{1x}$ is an $x$-directional gradient term defined as:

$$\boldsymbol{\nabla}_c^{1x} = \boldsymbol{\phi}\left(\frac{\mathbf{U}_c^0(t) - \mathbf{U}_{nei_W}^0(t)}{\mathbf{U}_{nei_E}^0(t) - \mathbf{U}_c^0(t)}\right)\left(\mathbf{U}_c^0(t) - \mathbf{U}_{nei_W}^0(t)\right) \qquad (10)$$

$\boldsymbol{\phi}$ is a *slope limiter function* that is applied component-wise and can lead to the FV1 solver by setting $\boldsymbol{\phi}(r) = 0$. The choice for $\boldsymbol{\phi}$ is discussed and justified later in Sec. 2.2.3. Similarly, the limits $\mathbf{U}_p^\pm$ at the other three face-centred nodes $p$ can be reconstructed ($p = W, N, S$). Achieving these MUSCL reconstructions for all four limits $\mathbf{U}_p^\pm$ in an element $Q_c$ requires a wider calculation stencil that needs further access to the neighbours' neighbour piecewise-constant data.

A different approach is applied to form the dummy limits $\mathbf{U}_p^{\pm,*}$ and $z_p^*$ ($p = E, W, N$ and $S$), based on an intermediate involvement of the free-surface elevation $h + z$. This involvement is necessary to ensure well-balancedness given the inherent involvement of the slope limiter in the FV2 scheme (Liang and Marche, 2009). For example, at $E$, the starting point is to construct two dummy topography limits $z_E^\pm$, via Eq. (9), as $z_E^\pm = (h+z)_E^\pm - h_E^\pm$, from which $z_E^* = \max(z_E^-, z_E^+)$ is computed to ensure the continuity property. Next, dummy water depth limits $h_E^{\pm,*}$ are reconstructed as $h_E^{\pm,*} = \max(0, (h+z)_E^\pm - z_E^*)$, through which the discharge limits become $(hu)_E^{\pm,*} = h_E^{\pm,*}u_E^\pm$ and $(hv)_E^{\pm,*} = h_E^{\pm,*} v_E^{\pm}$ on wet elements. The adjustment $z_E^* = z_E^* - \max\{0, -[(h+z)_E^- - z_E^*]\}$ is also applied to preserve well-balancedness when a motionless body of water is blocked by a high wall. The dummy revisions for $\bar{h}_c^{0x}$, $\bar{z}_c^{1x}$, $\bar{h}_c^{0y}$ and $\bar{z}_c^{1y}$ remain the same as in Eq. (5). These positivity preserving revisions are valid with Eq. (8) as long as the Courant number $Cr$ does not exceed 0.5, in order to preserve water depth-positivity over a time-step update (Kesserwani and Liang, 2012b).

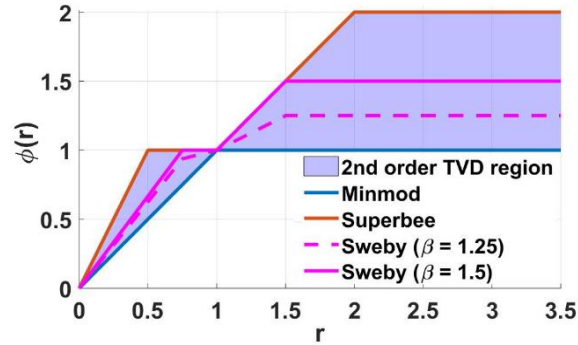

**Fig. 2.** Second-order TVD region proposed in Sweby (1984) for the choice of a valid slope limiter within a MUSCL-FV2 scheme.

### 2.2.3. Friction source term and choice for the limiter

The friction effects are integrated by following the same procedure as the DG2 model. It suffices to reapply the same treatment to update discharges within the average coefficient. However, the slope limiting process is substantially different from the MUSCL-FV2 scheme where the *limiter function* $\boldsymbol{\phi}(\boldsymbol{r})$ becomes a core component to both prevent spurious oscillations near sharp solution gradients and establish second-order accuracy. The choice for $\boldsymbol{\phi}(\boldsymbol{r})$ is usually made such that it is within the 2nd order TVD region proposed in Sweby (1984) which is bounded by two limiters, $minmod$ and $superbee$, that form the lower and upper limit respectively, as shown in Fig. 2. These limiters are not practically recommended (Sanders and Bradford, 2006): the $minmod$ limiter does not use slope amplification and thus yields the most solution smearing, but this pays off with more stability in particular at wet-dry fronts, as the two-argument $minmod$ limiter zeroes opposite sign slopes. In contrast, the $superbee$ limiter involves maximum slope amplifications, but it can over-amplify the slopes and cause severe instabilities at wet-dry fronts. A sensible choice for a limiter function in flood modelling is one that is strictly below the $superbee$ limiter bound and is fairly higher than the *minmod* limiter bound as well. Such a choice includes the generalised *Sweby symmetric limiter* with the $\beta$ parameter ranging $1 < \beta < 2$ (An and Yu, 2014) and the *Barth-Jespersen limiter* (Berger et al., 2005) among many others. In this paper, the *Sweby limiter* is selected subject to using $\beta = 1.25$ instead of the commonly reported value of 1.5, in order to make this limiter lower than the upper bound (see Fig. 2). This choice for $\beta$ is also supported by a sensitivity analysis (not presented here) conducted over the synthetic test case in Sec. 2.3, which has confirmed that the condition $\beta < 1.3$ is necessary to preserve the stability of the MUSCL-FV2 solver around wet-dry fronts.

**2.3. Analytical assessments of model conservation properties**

The synthetic test of curved oscillatory flow in a parabolic bowl presented in (Kesserwani and Wang, 2014) is reconsidered, for which both MUSCL-FV2 and DG2 models could equally reproduce the analytical solution at grid resolutions of 2 m × 2 m and 10 m × 10 m, respectively. Both models have also been demonstrated to deliver second-order grid convergence when simulating only one cycle of wetting and drying, with DG2 producing better quality velocity outputs. The aforementioned test is revisited in order to explore the numerical mass and energy conservation properties of the DG2 model compared to the FV-based models over much longer time evolution and for relatively coarse grid modelling. In this test and also later in Sec. 3, all models are run on their maximum allowable *Cr* number.

The curved oscillatory flow test case (Thacker, 1981) is applied with the same test setup used by other investigators (Bunya et al., 2009; Ern et al., 2008; Kesserwani and Wang, 2014). Initially ($t$ = 0 s), the discharges are zero and the water level follows a concave profile, for which the wet-dry frontline is at its narrowest extent (see analytical profile in Fig. 3b). This state of the flow is termed *maximally dry profile* and should be retrieved after a period cycle $\tau = 1756.2$ s, as the oscillatory flow runs on a frictionless domain. The opposite situation is expected to occur after half-a-period cycle, at $t = \tau/2$, leading to the *maximally wet profile* characterised by a convex profile for which the wet-dry frontline is at its widest extent (see analytical profile in Fig. 3a). A grid with a resolution of 40 m × 40 m is applied to discretise the 2D spatial domain where the DG2, MUSCL-FV2 and FV1 models are run up to 30-period cycles ($t = 30\tau$). Simulation results are presented for two DG2 variants with LL (DG2-LL) and with no limiting (DG2-NL), in terms of water level predictions to the maximally dry and maximally wet profiles after 7-period cycles, as shown in Fig. 3.

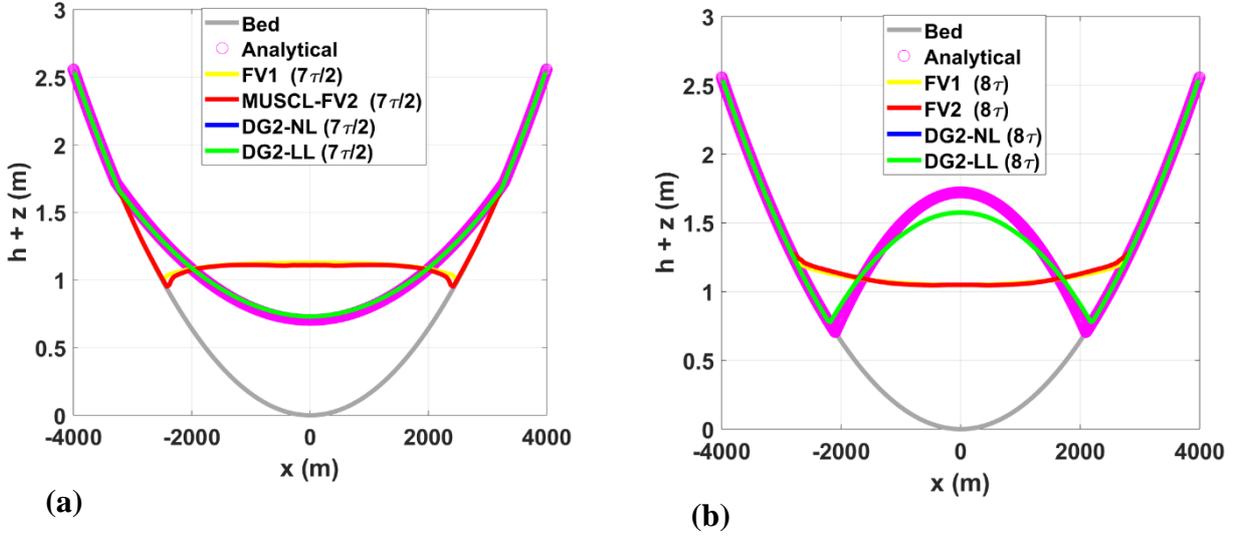

**Fig. 3.** Water level predicted by the DG2, FV2 and FV1 models extracted along the $x$-directional centreline after 7-period cycles, when the domain is: (a) maximally wet at $t = 7\tau/2$, and (b) maximally dry at $t = 8\tau$.

The results in Fig. 3, after $t = 7\tau$, are particularly useful to examine the extent to which the DG2-LL, DG2-NL, MUSCL-FV2 and FV1 models are impacted by the growth of numerical diffusion, and how this impact eventually manifests within the model predictions. Both DG2 model predictions are found identical for this test, remaining close to the analytical profiles, whereas MUSCL-FV2 and FV1 model predictions are relatively poorer. FV1 and MUSCL-FV2 models lead to a narrower wetting extent and an over-prediction of the water level in the wet zone for the maximally wet profile (Fig. 3a). For the maximally dry profile (Fig. 3b), the FV1 and MUSCL-FV2 models lead to a wider wetting extent and an under-prediction of water level in the wet zone.

To quantify the impact of the accumulated numerical diffusion on the conservation properties of the DG2-LL, DG2-NL, MUSCL-FV2 and FV1 solvers over the 30-period cycle ($t = 30\tau$), mass error and total energy are measured. This impact is also quantified by considering twice-finer grids (i.e. resolution of 20 m) to gain further insight into the different model response to resolution coarsening. As the flow oscillates in an enclosed system, there is an exchange between kinetic and potential energy, but the total energy should be conserved as there is no friction. The 2D domain-integrated total mass $M$ is calculated as:

$$M(t) = \int_{-4000}^{+4000} \int_{-4000}^{+4000} h(x,y,t) \, dxdy \tag{11}$$

The mass error (i.e. the difference between the initial and current total mass) is calculated over time and normalised to the initial mass $M_0 = M(0)$. Also, the 2D domain-integrated total energy $E$ is calculated using the following relationship (McRae, 2015):

$$E(t) = \int_{-4000}^{+4000} \int_{-4000}^{+4000} (\frac{1}{2} h(x,y,t) \, (u(x,y,t)^2 + v(x,y,t)^2) + \frac{1}{2} g((h(x,y,t) + z(x,y))^2 - z(x,y)^2)) \, dxdy \tag{12}$$

and is then normalised to the initial energy $E_0 = E(0)$. While a perfectly energy-conservative scheme yields 1 for the normalised total energy, a value below 1 is indicative of the impact of accumulated numerical diffusion as time evolves.

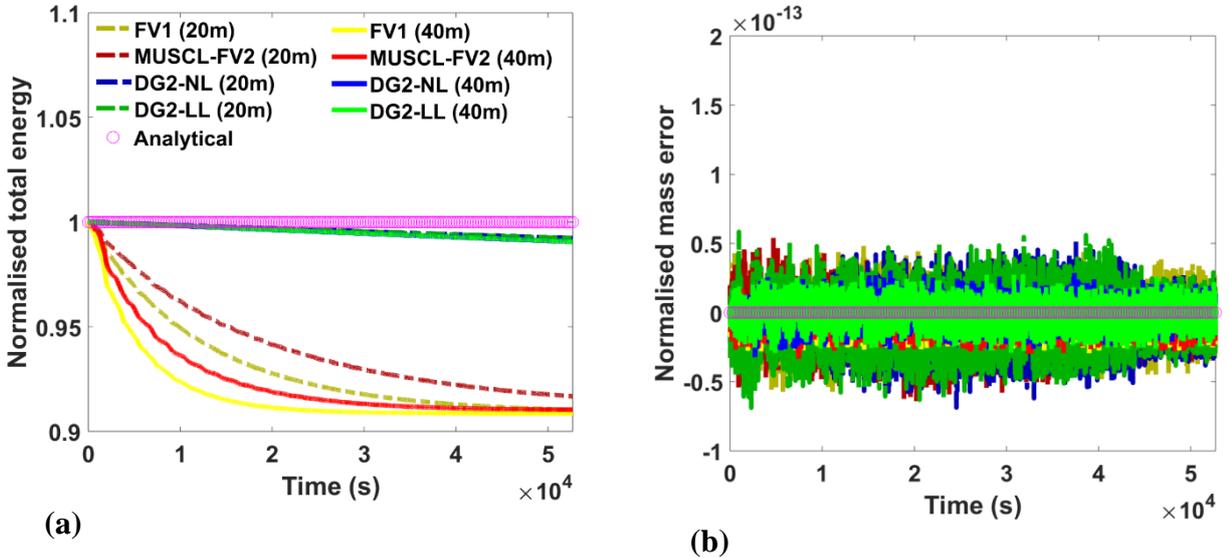

**Fig. 4.** Time histories of the 2D domain-integrated: (a) normalised total energy and (b) mass error, for the grids with a resolution of 20 m and 40 m, respectively, and over a simulation time of $t = 30\tau$.

Fig. 4 shows the time histories of the normalised total energy and the mass errors recorded from the DG2-LL, DG2-NL, MUSCL-FV2 and FV1 outputs for two investigated grid resolutions (20 m and 40 m). As expected, the net effect of numerical diffusion in the DG2 variants, MUSCL-FV2 and FV1 solvers induces a gradual decay in the total energy (Fig. 4a). Even on the finer grid, the FV-based models are observed to dissipate energy quite rapidly, around 9% of the initial energy after 30 period cycles. This dissipation of initial energy is observed earlier with the coarser grid, which indicates that FV-based models are more prone to losing energy as grid resolution is coarsened. In contrast, both DG2-LL and DG2-NL variants exhibit better energy conservation properties, with around 1% loss of the initial total energy despite grid resolution coarsening. This is because the DG2 method evolves its local slope coefficients based on a local discrete translation of the conservative model equations (Eq. 4b and Eq. 4c) involving inter-elemental Riemann fluxes. Apart from the discrepancies observed in terms of numerical energy conservation, all models yield a zero relative mass error (Fig. 4b), suggesting that they are perfectly mass-conservative.

Overall, this test suggests that the DG2-NL variant can be as good as the DG2-LL when modelling (shockless) shallow water flows with wetting and drying. The DG2 solver is equally mass conservative as the FV-based solvers, but is more energy conservative due to its superior resistance to numerical dissipation. This property seems to make a DG2-based model better suited for hydraulic modelling over a long timescale and/or on coarser grids.

**Table 1**. List of the industrial models considered for the comparison against the DG2 models.

| Model name | Supplier | Numerical scheme | Mesh type |
|---|---|---|---|
| TUFLOW FV1 | BMT-WBM | FV1 with $1^{st}$ order time integration | Flexible mesh |
| TUFLOW FV2 | BMT-WBM | MUSCL-FV2 with $2^{nd}$ order time integration | Flexible mesh |
| TUFLOW HPC | BMT-WBM | MUSCL-FV2 with $4^{th}$ order time integration | Square grid |
| Infoworks ICM | Innovyze | FV1 with $1^{st}$ order time integration | Flexible mesh |

## 3. Comparison against industrial flood model outputs

The assessment of the DG2 model is based on five industrially-relevant test cases recommended by the UK Environment Agency (Neelz, S., Pender, 2010; Neelz and Pender, 2013), which have already been studied with various industrial flood models. Comparative analyses are made with respect to available outputs from four, well-established industrial FV-based flood models (Table 1), all adopting the Roe Riemann solver within an FV1 scheme or an FV2 scheme based on linear (MUSCL) reconstructions with TVD slope limiting (BMT-WBM, 2018, 2016; Huxley et al., 2017; Innovyze, 2011; Jamieson et al., 2012; Lhomme et al., 2010; Neelz and Pender, 2009). The test cases have been categorised as to whether the flood flow propagation is gradual (Froude number < 1) or rapid (Froude number ≥ 1), with their specific description parameters shown in Table 2.

For each test case, DG2 model simulations are run on the same grid resolution as reported for the industrial model outputs (Table 2). DG2-LL and DG2-NL simulations are performed with a dual purpose: (a) to further diagnose the utility of the DG2-NL alternative for simulating flood-like flows; and (b) to uncover trade-offs related to the standard DG2-LL alternative. On a twice-coarser grid, the best performer DG2 variant (DG2-NL is selected by default if both perform equally well) is rerun alongside a MUSCL-FV2 run to also assess the different model responses to doubling the resolution.

**Table 2.** Selected benchmark test cases with their specific parameters (Neelz, S., Pender, 2010; Neelz and Pender, 2013).

| Benchmark test cases | Finest DEM resolution (m) | Reported resolution (m) | $n_M$ | Output time $T$ |
|---|---|---|---|---|
| **Gradual flood propagation (Sec. 3.1)** | | | | |
| Flooding and drying cycle over a sloping topography (Sec. 3.1.1) | 2 | 10 | 0.03 | 20 hrs |
| Slow filling of multiple ponds (Sec. 3.1.2) | 2 | 20 | 0.03 | 48 hrs |
| **Rapid flood propagation (Sec. 3.2)** | | | | |
| Momentum conservation over an obstruction (Sec. 3.2.1) | 2 | 5 | 0.01 | 900 sec |
| Torrential flooding over a rural catchment (Sec. 3.2.2) | 10 | 50 | 0.04 | 30 hrs |
| Dam break over an oblique building (Sec. 3.2.3) | 0.05 | 0.1 | 0.01 | 120 sec |

MUSCL-FV2 simulations are also performed on the finest DEM resolution available (Table 2) for the first four test cases to form a *reference solution*, to which all flood model outputs are compared qualitatively and measured quantitatively by two statistic indices: the R-squared ($R^2$) coefficient, or coefficient of determination, and the $L^1$-norm error. These indices are evaluated for the water level and/or velocity time series recorded at the sampling points specified for each test case. Their expressions are:

$$R^2 = \left[\frac{\sum_{t=1}^{T}(V_t^R - \bar{V}^R)(V_t^M - \bar{V}^M)}{\sqrt{\sum_{t=1}^{T}(V_t^R - \bar{V}^R)^2 \sum_{t=1}^{T}(V_t^M - \bar{V}^M)^2}}\right]^2 \qquad (13)$$

$$\text{L}^1\text{-norm error} = \frac{1}{N_s}(\textstyle\sum_{t=1}^{T}|V_t^R - V_t^M|) \tag{14}$$

where $t$ denotes the current time, $T$ the simulation output time (Table 2). The values $V_t^R$ and $V_t^M$ refer to an entry from the reference and modelled time series at time $t$, respectively; $\bar{V}^M$ and $\bar{V}^R$ represent time-averaged mean values; and $N_s$ is the size of the time series. The $R^2$ coefficient takes values between 0 and 1, indicating stronger correlation to the reference solution as it gets closer to 1. The $R^2$ coefficient is useful to quantify the resemblance between the predicted solution and the reference solution (e.g. similarity in hydrograph shapes), whereas the $L^1$-norm error provides an estimate of the average deviation of a model prediction from a reference solution (e.g. closeness of the predicted data to the reference data). Runtime costs are also analysed between MUSCL-FV2 and the DG2 variants for the same resolution of the industrial model outputs and for the grid with twice-coarser resolution (Table 2).

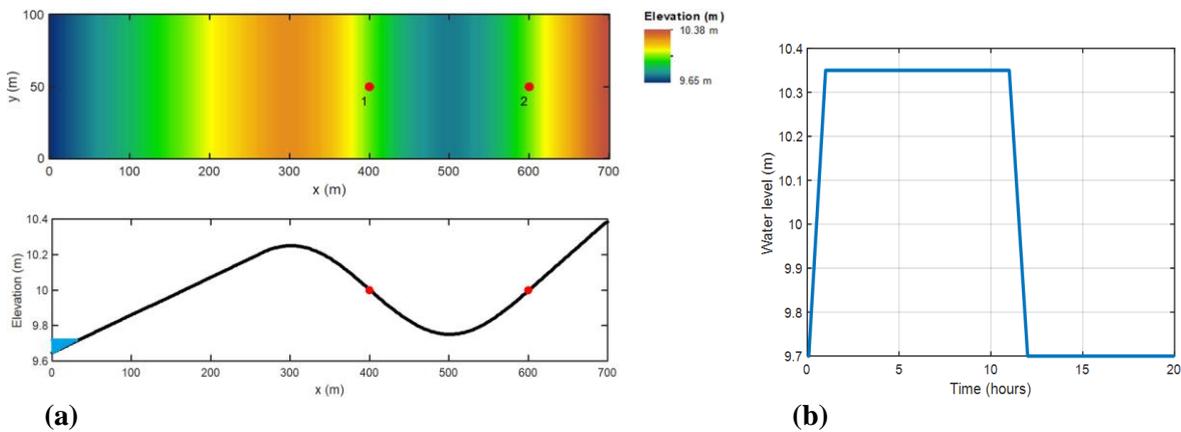

(a)          (b)

**Fig. 5**. Flooding and drying cycle over a sloping topography: (a) plan view (upper panel) and longitudinal profile (lower panel) of the terrain including the location of the sampling points (red dots). Blue shaded area in the lower panel indicates the initial water volume in the domain; and, (b) inflow water level hydrograph imposed at the west boundary.

## 3.1. Gradual flood propagation

### 3.1.1. Flooding and drying cycle over a sloping topography

This test is often used to assess the ability of a flood model to reproduce a slow flooding and drying cycle occurring over a long duration. It involves a unidirectional flow moving over a 2D sloping topography ($n_M = 0.03$). The topography is initially dry except for a small water volume ponding near the west boundary with water level of 9.7 m (Fig. 5a, lower panel). The water inflow (Fig. 5b) at the west boundary pushes the surface flow upslope and then downslope to accumulate in the pond (Fig. 5a, from $x = 300$ m to $x = 700$ m). Over time, the accumulated water reaches the level 10.35 m, at which the whole topography becomes fully inundated. The water remains at this level until $t = 11$ hrs and then starts to recede towards the west boundary. Eventually, the topography dries out except in the pond, where the water level reduces to about 10.25 m. The time histories of the water level were recorded at the two sampling points shown in Fig. 5, and are presented in Fig. 6.

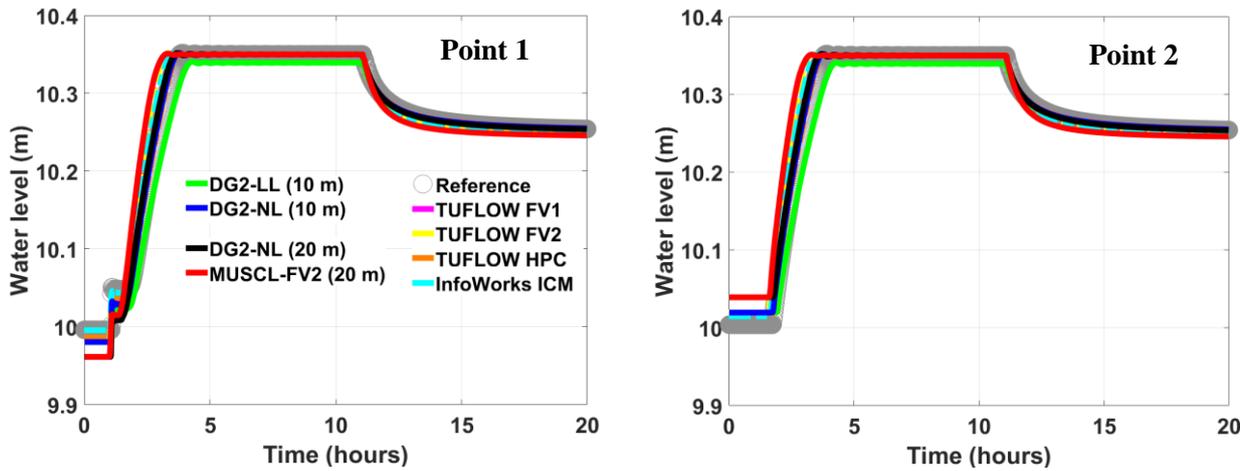

**Fig. 6.** Flooding and drying cycle over a sloping topography. Water level histories predicted by DG2-LL and DG2-NL at 10 m resolution, DG2-NL and MUSCL-FV2 models at 20 m resolution, the industrial models at 10 m resolution and the reference solution at 2 m resolution.

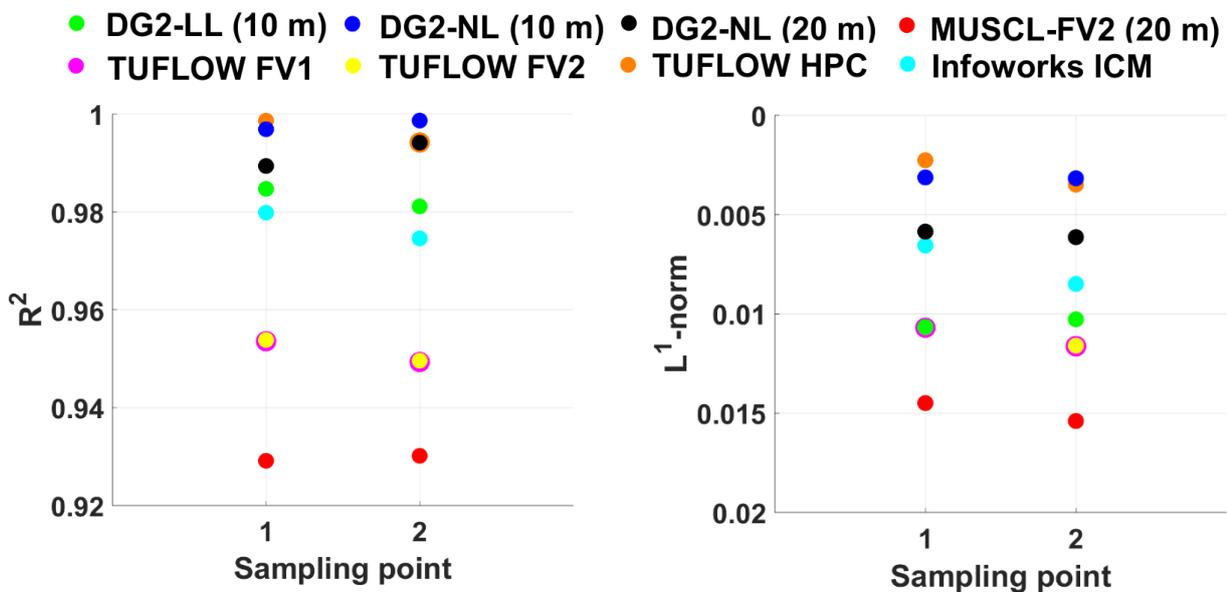

**Fig. 7.** Flooding and drying cycle over a sloping topography. Difference among the flood model predictions relative to the reference solution: $R^2$ coefficients (left) and $L^1$-norm error (right).

On the grid with 10 m resolution, DG2-LL provides water level predictions that are outside the range of the water levels predicted by the industrial models. This suggests that local limiting with DG2 could still impact the DG2 flood model predictions, in particular during the flooding stage when the inflow is still feeding into the domain. In contrast, DG2-NL produces predictions that are in the range of the industrial model predictions and in closer agreement with the reference solution (at a 2 m resolution), even on a grid at 20 m resolution. The plots of the $R^2$ coefficient and the $L^1$-norm error in Fig. 7 confirm that DG2-NL excels in producing the most similar and the closest water level histories in relation to the reference solutions, even at 20 m resolution. DG2-LL at 10 m resolution, although providing similar shape as the reference solution, leads to predictions that are relatively more deviated than DG2-NL at 20 m resolution. This test points out the potential of using the DG2-NL variant to model gradual flooding and drying over a long period, and reproduce the industrial model predictions at twice-as-coarse grid resolution.

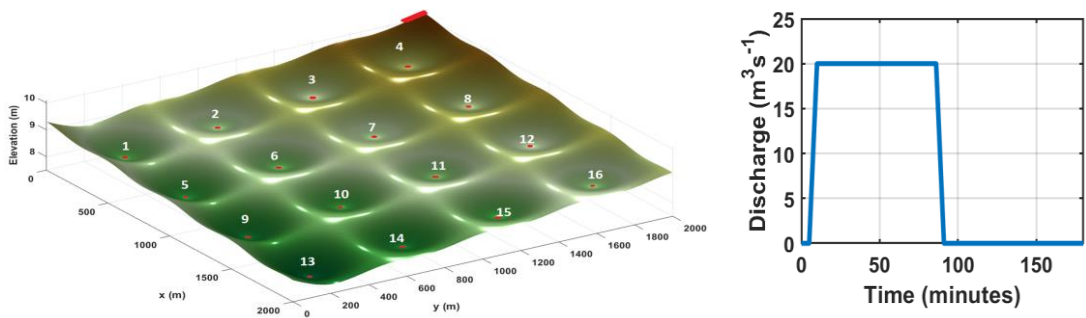

**Fig. 8.** Slow filling of multiple ponds: the left panel shows the 2D terrain including 16 ponds, the location of the breach (red line, top corner) and the sampling points (red dots) at the middle of each pond; the right panel illustrates the inflow hydrograph imposed at the breach.

### 3.1.2. Slow filling of multiple ponds

This test involves a long-duration flooding over a down-sloping topography ($n_M = 0.03$) with 16 smooth ponds (Fig. 8). The surface slope in the *y*-direction is about twice of that in the *x*-direction. The highest elevation of the topography is at (0 m, 2000 m), whereas the lowest one is at (2000 m, 0 m). An inflow discharge with a peak of 20 m$^3$ s$^{-1}$ (Fig. 8) flows from a 100 m breach opening (red line, Fig. 8) over the initially dry terrain. When the inflow is active, the propagating flood gradually fills seven ponds near the breach (covering the points 4, 3, 2 and 1, and the points 8, 7 and 6 shown in Fig. 8) until the inflow empties at $t = 1.5$ hrs. Then, driven by gravity, the flow slowly propagates into farther ponds covering the points 5, 10, 11 and 12, until the simulation terminates at $t = 48$ hrs. As the ponds at points 9, 13, 14, 15 and 16 remain dry during the simulation, the time histories of water level are recorded only at the remaining eleven points.

Fig. 9 contains the time histories of the water levels in which no difference is identified between the predictions made by DG2-NL and DG2-LL at 20 m resolution, both delivering the best agreement with the reference solution (at a 2 m resolution). Near the breach (i.e. at points 4, 3, 2 and 1, and points 8, 7 and 6), all DG2 model predictions are found to be closer to the range of predictions made by the industrial models, as opposed to the MUSCL-FV2 predictions at 40 m resolution, which are slightly out of the range for $t > 1.5$ hrs. Further away from the breach (i.e. at points 5, 10, 11 and 12), discrepancies start to emerge, and the MUSCL-FV2 at 40 m resolution predicts the highest water levels amongst all models, whereas the DG2 variants at 20 m resolution predict the lowest. The DG2-NL at 40 m resolution is seen to deliver water levels and arrival times that are relatively closer to the range of predictions made by the industrial models than those obtained with MUSCL-FV2.

By analysing the $R^2$ coefficient plots in Fig. 10, the largest difference in the water histories patterns is identified at the points 5, 11 and 12, where the DG2 variants at 20 m resolution offer the best agreement. Notably, DG2-NL at 40 m resolution reproduces a shape that is compatible with the reference solution ($R^2 > 0.9$) despite leading to relatively higher deviations via the $L^1$-norm errors. At the points located close to the breach (i.e. 4, 3, 2, 1, 8, 7 and 6), DG2-NL at 40 m resolution leads to $R^2$ coefficients and $L^1$-norm errors that are very close to those of the industrial models and DG2 variants at 20 m resolution, which is in line with the findings in Sec. 3.1.1.

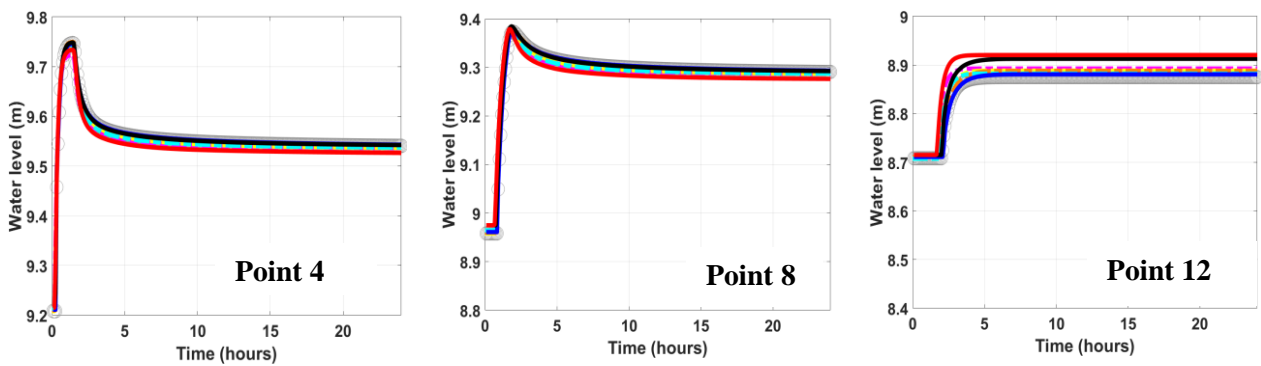

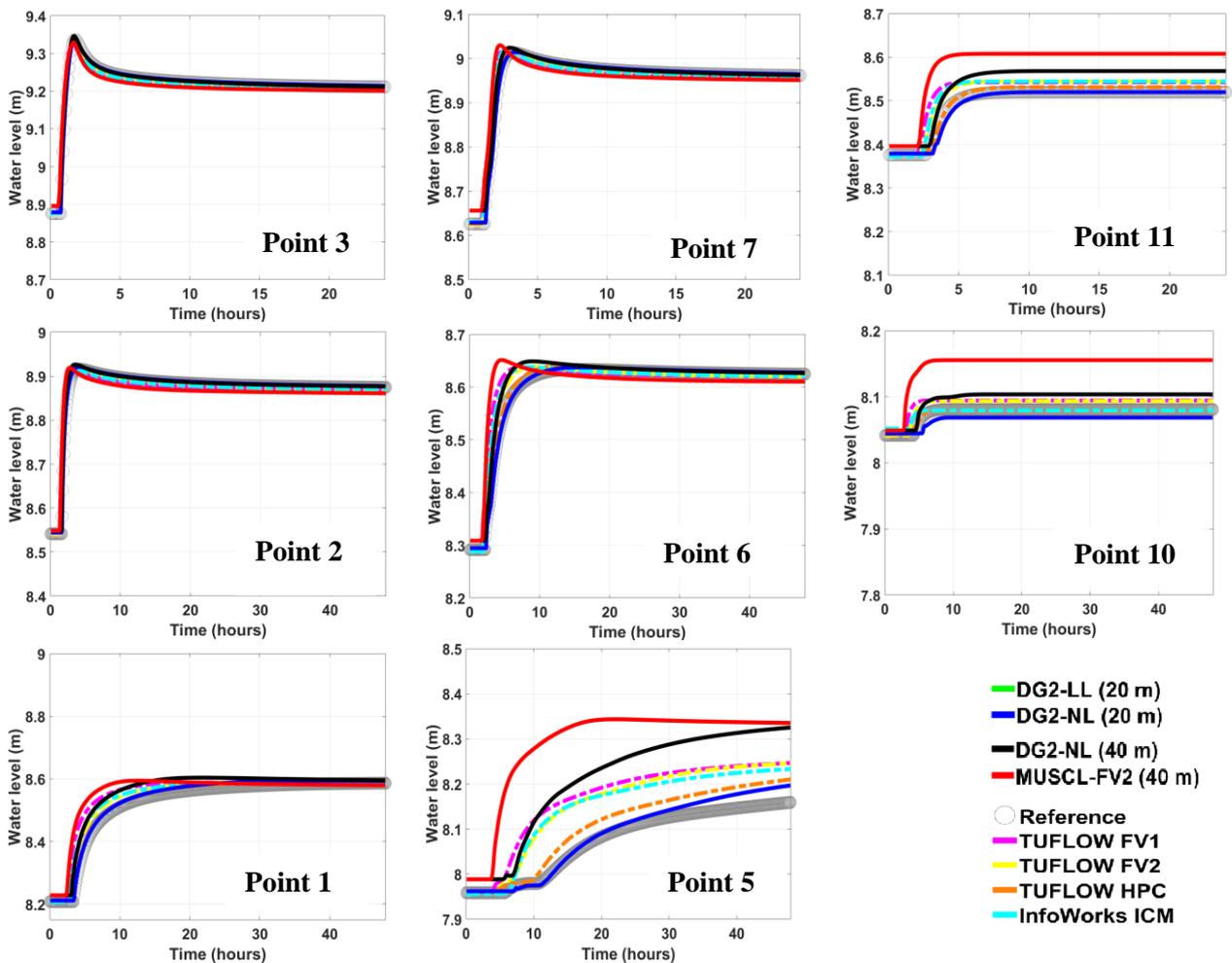

**Fig. 9.** Slow filling of multiple ponds. Water level histories recorded from the DG2-LL and DG2-NL simulations at 20 m resolution, DG2-NL and MUSCL-FV2 models at 40 m resolution, the industrial models at 20 m resolution and the reference solution at 2 m resolution.

Overall, this test shows that a DG2 alternative allows more accurate capturing of hydrographs over long time evolution even when taken at the same resolution as the industrial models. With a twice-as-coarse resolution, the DG2 alternative tends to produce deviated hydrographs as the location of the gauges become far away from the breach (i.e. sufficiently away from the flooding source for the flood flow propagation to be solely driven by topographic connectivities and decelerated by frictional forces). Still, compared to MUSCL-FV2, the DG2 model at a twice-as-coarse resolution can produce hydrographs that better correlate with those produced by the industrial models.

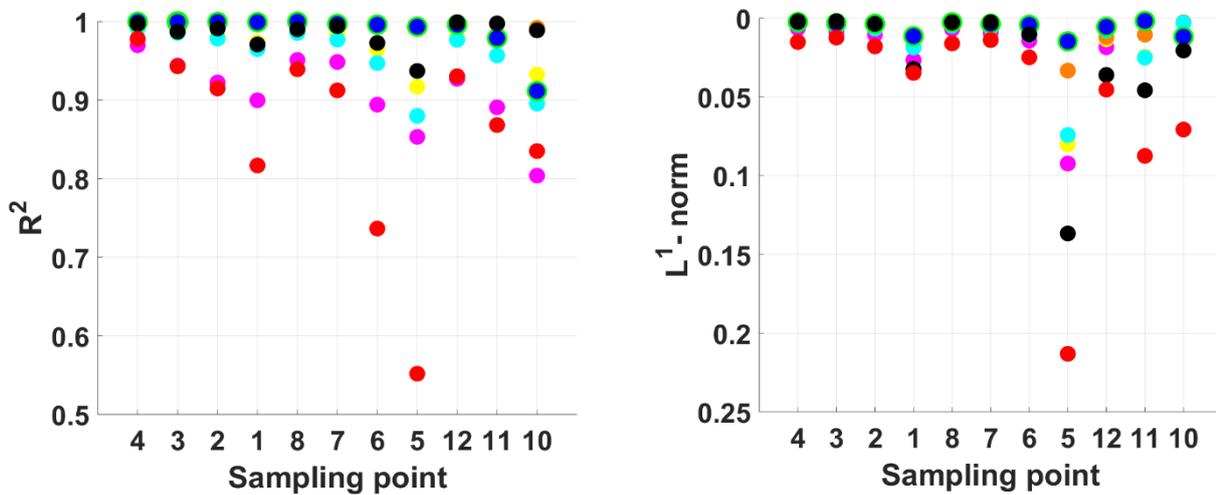

**Fig. 10.** Slow filling of multiple ponds. Difference among the flood model predictions relative to the reference solution: $R^2$ coefficients (left) and $L^1$-norm error (right).

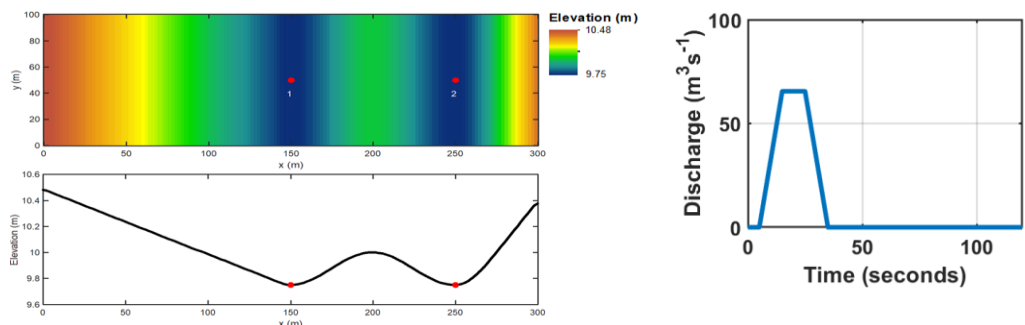

**Fig. 11.** Momentum conservation over an obstruction. Plan view and longitudinal profile of the topography with location of sampling points marked by red dots (left panel); inflow discharge hydrograph imposed at the west boundary (right panel).

### 3.2. Rapid flood propagation

#### 3.2.1. Momentum conservation over an obstruction

This test involves a rapid unidirectional flow, filling two initially dry ponds separated by a hump at $x = 200$ m over a short duration. The 2D topography (Fig. 11) has a smoother surface allowing faster flow acceleration ($n_M = 0.01$ s m$^{1/3}$). The inflow has a peak discharge of 65.5 m$^3$ s$^{-1}$ (Fig. 11) and accelerates downslope from the west boundary to fill the first pond (Fig. 11, from $x = 100$ m to $x = 200$ m). While the first pond is large enough to hold the total inflow volume, models solving the full SWE tend to predict a small volume of water overtopping the hump to fill the second pond (Fig. 11, from $x = 200$ m to $x = 300$ m). Once all ponds are filled, the water within each pond oscillates until it completely settles. At sampling points 1 and 2 (marked in Fig. 11), the time histories of the water level and velocity (at point 1 only) are recorded and compared in Fig. 12 against those of the industrial models and the reference solution (at a 2 m resolution).

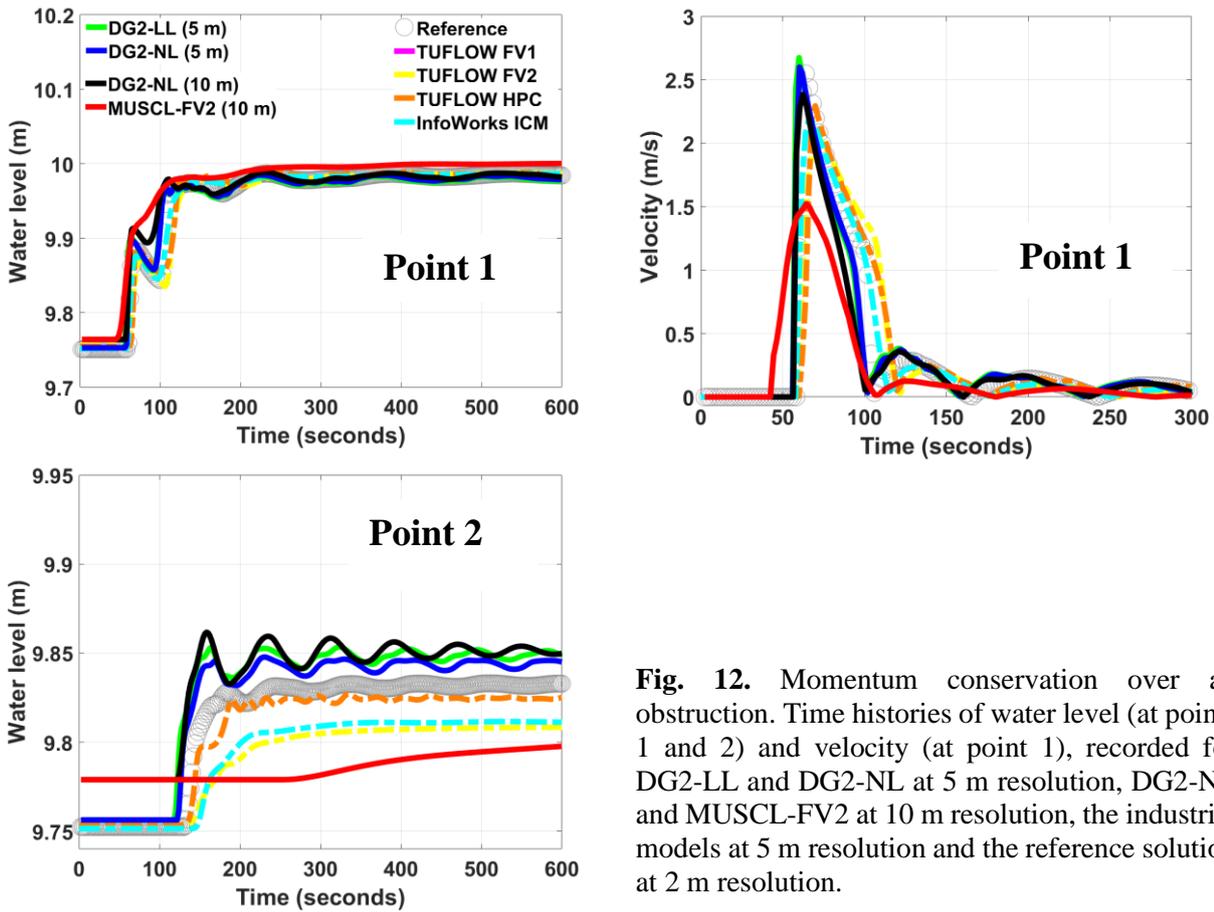

**Fig. 12.** Momentum conservation over an obstruction. Time histories of water level (at points 1 and 2) and velocity (at point 1), recorded for DG2-LL and DG2-NL at 5 m resolution, DG2-NL and MUSCL-FV2 at 10 m resolution, the industrial models at 5 m resolution and the reference solution at 2 m resolution.

For the water level histories at point 1, the DG2 variants at 5 m resolution predicted slightly earlier wave arrival and faster flow-rising rate compared to the industrial models. These DG2 variants are also the only models that are able to capture the undulating profile visible in the reference solution (Fig. 12, left part in the upper panel). In terms of velocity histories, all DG2 variants at 5 and 10 m resolution exhibit a sharper peak alongside a faster rate of recession, vanishing and occurrence than the industrial models, and result in a more consistent agreement with the reference solution. The velocity histories at point 1 therefore suggests that the DG2 variants at 5 and 10 m resolution induce faster water filling patterns compared to the FV-based industrial models and the reference solution. This observation is confirmed by the water level histories of the DG2 variants at point 2. Their predicted arrival times are earlier (compare with those of point 1), which leads to faster filling, better momentum capturing, and a more defined oscillatory flow pattern that manifests as distinct undulating shapes in DG2-related water level histories (Fig. 12, left part in the lower panel). These shapes are not seen in the water level histories made by the FV-based industrial models, although this is barely observable in the reference solution. No major differences are noted between the DG2-NL and DG2-LL predictions, but DG2-NL is in a slightly better agreement with the reference solution for this test (Fig. 12, left part in the lower panel; see also Fig. 13).

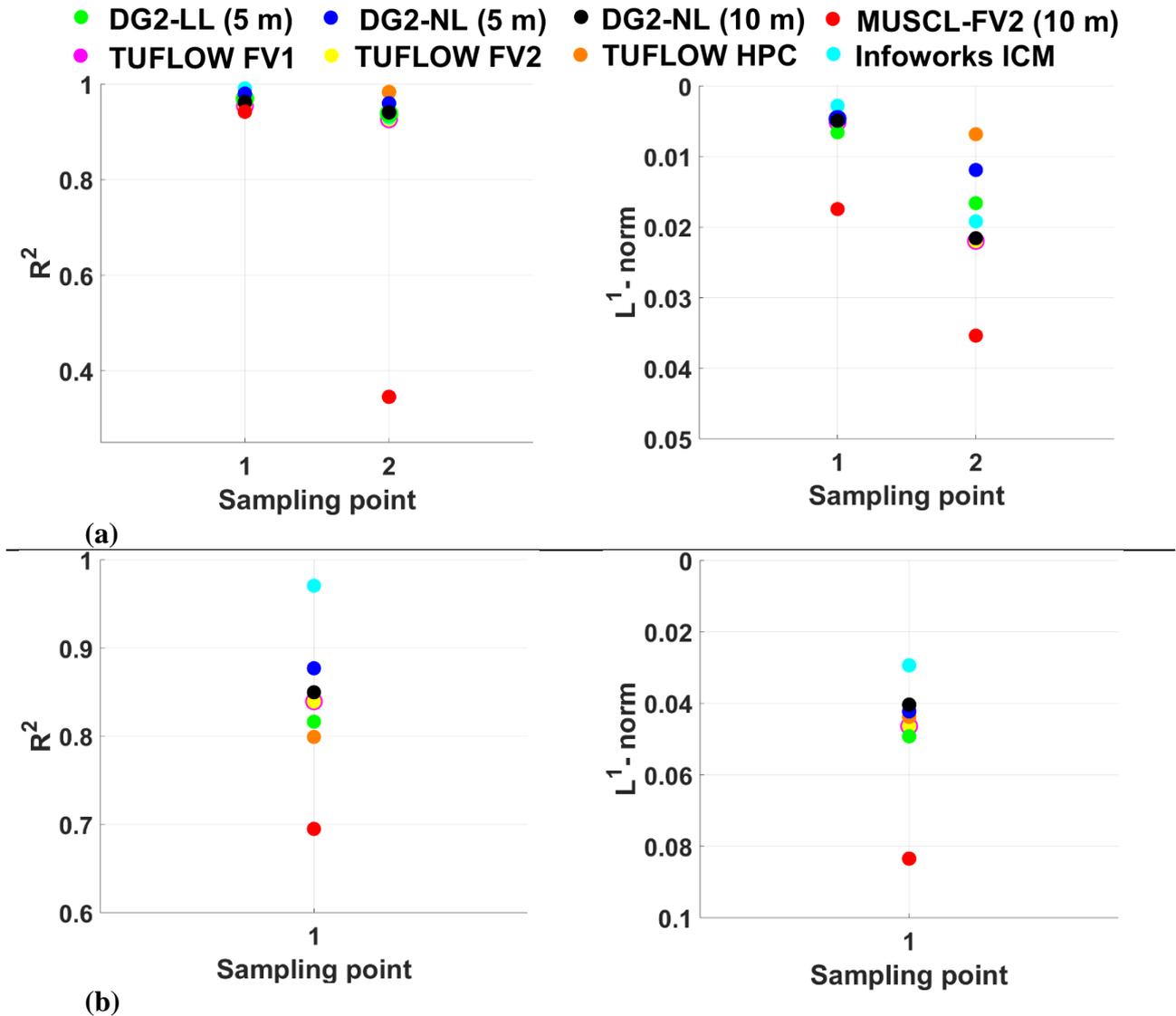

**Fig. 13.** Momentum conservation over an obstruction. Difference among the flood model predictions relative to the reference solution: $R^2$ coefficients (left) and $L^1$-norm errors (right); (a) water level and (b) velocity.

The analysis of the $R^2$ coefficient and $L^1$-norm error plots reinforces the DG2 variants' ability to predict profiles that are closer to the shape of the reference solution and within the range of deviation of the industrial model outputs. These metrics also allow to clearly distinguish the performance between the DG2 variants, suggesting that DG2-NL at 5 m resolution delivers better agreement with the reference solution, while still leading to competitive results at 10 m resolution. This test confirms (recall Sec. 3.2.2) the benefits of the DG2-NL alternative to also improve capturing of fast (shockless) flow, for example in flash flooding, where its better response to resolution coarsening is found to capture velocity variations better.

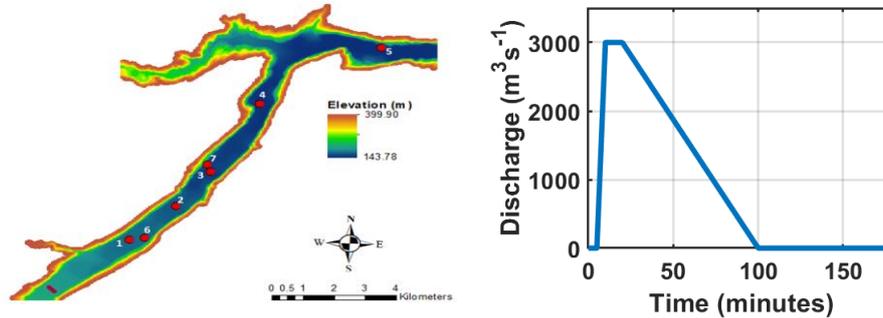

**Fig. 14.** Torrential flooding over a rural catchment. Left panel shows the 2D topography with the breach location (red line, southwest of the catchment) and the sampling points (red dots). The right panel shows inflow hydrograph imposed at the breach.

### 3.2.2. Torrential flooding over a rural catchment

This test considers a long-duration torrential flooding occurring over a down-sloping valley (Fig. 14) within a 17.0 km × 0.8 km rural catchment, with a naturally rugged ($n_M = 0.04$) and initially dry topography. Flooding inflow occurs from a hydrograph casting a dam breach located southwest of the valley with a peak of 3,000 m$^3$ s$^{-1}$ (Fig. 14). The breach has a 260-m wide opening, through which the flood rapidly advances to accumulate downstream on the east side of the valley (i.e. within the area located at sampling point 5, shown in Fig. 14). As the flood advances, it is expected to fill shallow ground depressions along the valley. This flooding is simulated until $t = 30$ hrs. The time histories of the water level and velocity, recorded at the sampling points (marked in Fig. 14), are shown in Fig. 15, along with the histories of the industrial models and the reference solution (at 10 m resolution).

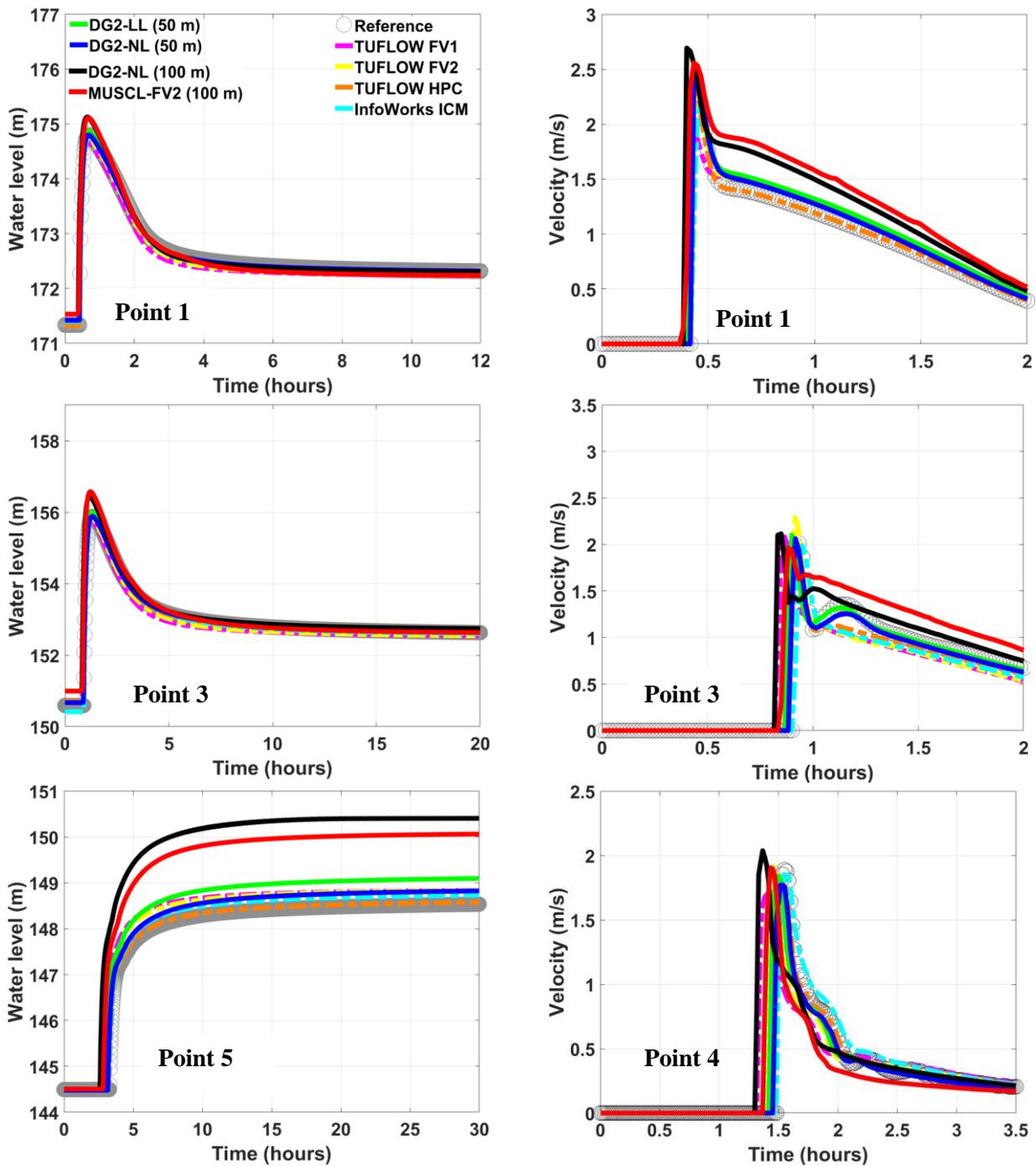

**Fig. 15.** Torrential flooding over a rural catchment. Time histories of the water level (left side) at point 1, point 3 and point 5, and of the velocity (right side) at point 1, point 3 and point 4, produced by the DG2-LL and DG2-NL at 50 m resolution, by the DG2-NL and MUSCL-FV2 models at 100 m resolution, against those produced by the industrial models at 50 m resolution and the reference solution at 10 m resolution.

Fig. 15 (right side) contains the velocity histories at point 1 and point 3 (during 2 hrs), located within the mainstream flow direction (point 1 close to the breach and point 3 by the middle of the valley), and point 4 (during 3.5 hrs) located at the downstream end of the valley. At point 1, results are analysed during the first 1.5 hrs when there is still momentum forcing from the inflow: DG2 and MUSCL-FV2 at 100 m resolution yield velocities that are deviated higher (thus faster flow) from those predicted by the other models at 50 m resolution and the reference solution, although still producing the same velocity patterns. At point 3, the DG2 variants at 50 m resolution capture the transient curving observed within the reference solution (~1-1.3 hrs), which is not reproduced by any of the industrial models. At point 4, the velocities predicted by DG2 and MUSCL-FV2 at 100 m resolution are less deviated from the predictions made at 50 m resolution by the DG2 variants and the industrial models. This convergence in performance seems to arise from the fact that the flow that passes point 4 after 1.5 hrs is relatively slower (i.e. entirely driven by gravity when inflow-forcing is no longer in place and decelerated by high friction effects). However, at point 4, none of the models is able to capture the undulating velocity patterns observed in the reference solution (~2-2.8 hrs), which indicates that finer than 50 m resolution is required for the DG2 variants to observe these velocity transients.

Fig. 15 (left side) shows the water level histories at the same points 1 and 3, and then at point 5 located farthest from the flooding source. At points 1 and 3, no major discrepancies are observed between the model predictions, at both 50 m and 100 m resolution). All models exhibit very similar water level profiles and closely match the reference solution. At point 5 where the flow arrives around 4.5 hrs of flooding, DG2 and MUSCL-FV2 at 100 m resolution predict patterns for the water level histories that are entirely consistent with those of the industrial models and DG2 variants at 50 m resolution. However, at 100 m resolution, DG2 and MUSCL-FV2 clearly present the most deviated histories from the reference solution and from the predictions made by the DG2 variants and the industrial models at 50 m resolution. Notably, the 100 m resolution models predicted faster flow arrival, (recall Fig. 15, right side, points 1 and 3), to record a hydrograph for longer than 24 hrs. This finding again reinforces (recall Sec. 3.1.3) that a DG2 alternative at twice-coarser resolution would increasingly yield deviated water level hydrographs as the location of the gauges for analysis become increasingly away from the flooding source.

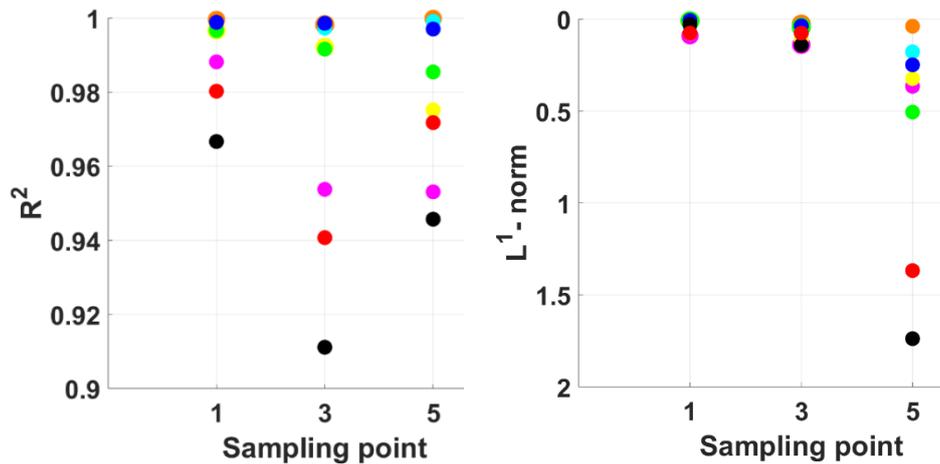

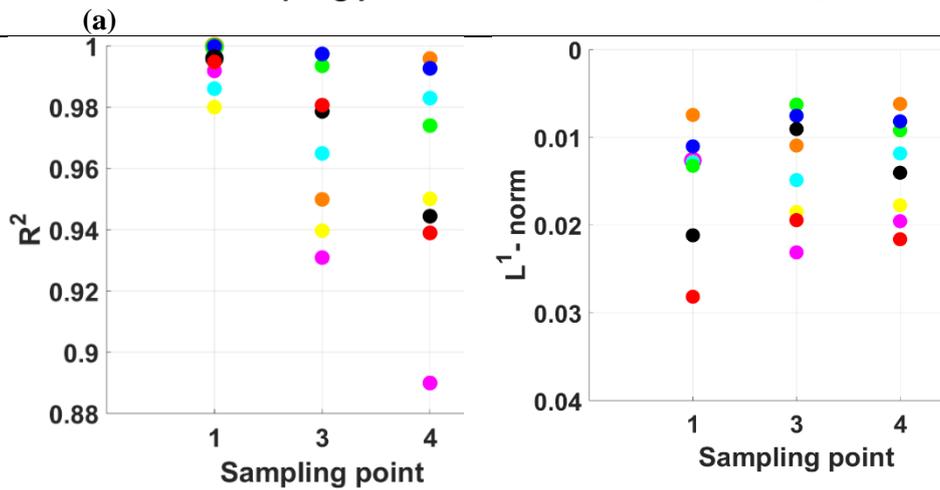

**Fig. 16.** Torrential flooding over a rural catchment. Difference among the flood model predictions relative to the reference solution: $R^2$ coefficients (left) and $L^1$-norm errors (right); (a) water level and (b) velocity.

The observations discussed previously can be further supported by analysing the plots of the $R^2$ coefficients and $L^1$-norm errors in Fig. 16 at the respective sampling points. The $R^2$ coefficients for the water level and velocity are greater than 0.88, which indicates strong similarities with the hydrograph patterns of the reference solution. The $L^1$-norm error for water level predictions (Fig. 16a), is relatively bigger at point 5 for the models run at 100 m resolution. This indicates that resolution coarsening produces hydrographs that are fairly different than those obtained with the models run at 50 m resolution. In terms of velocity predictions (Fig. 16b), $L^1$-norm errors indicate that all model predictions at 50 and 100 m resolution yield a relatively good fit to the reference solution, and that the DG2 variants are among the best performers. Overall, this test shows that a DG2-NL alternative for coarse resolution modelling ($\geq$ 50 m resolution) pays off with a more accurate capturing of the velocity transients occurring in rapid flooding flows over a real terrain.

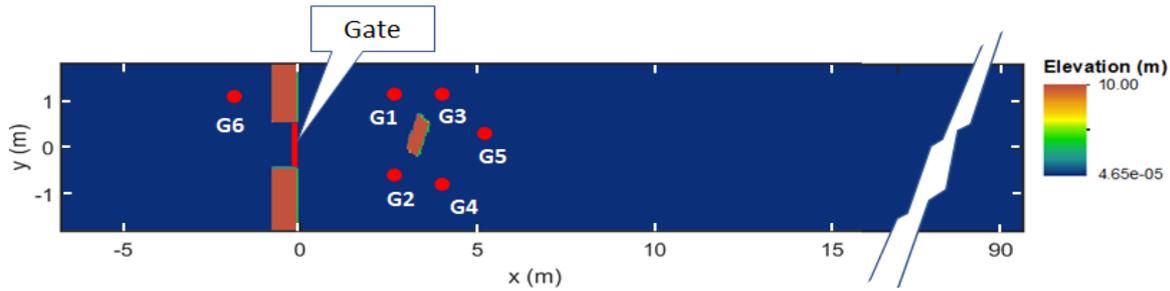

**Fig. 17**: Dam-break over an oblique building. Plan view of the spatial domain and the location of 6 sampling points (red dots).

### 3.2.3. Dam-break over an oblique building

This test case considers an abrupt wave propagation involving the reflection of multiple shock waves, the formation of a hydraulic jump, and the occurrence of small-scale eddies near wake zones. These phenomena are induced by the release of a dam-break wave over an initially wet floodplain to interact with an oblique building structure and solid walls enclosing the floodplain. The test case domain, shown in Fig. 17, is made of a long flume with a smooth bed ($n_M = 0.01$), and upstream walls with a gate initially separating a water body of height 0.4 m and forming a water height of 0.02 m. As the gate is swiftly opened, a shock wave rapidly propagates and collides with the building to form a reflected hydraulic jump (i.e. within the neighbourhood of G2, Fig. 17) and later splits into two shock waves that move in different directions (i.e. one towards G3 and one towards G4, Fig. 17). Downstream of the building, a wake zone emerges and surrounded by recurrent wave crossings, while producing small wake eddies (i.e. within the neighbourhood of G5, Fig. 17). 2D snapshots of these flow patterns can be found in Ginting (2019).

A grid with a cell size of 0.1 m is used to discretise the whole flume, including the oblique building and the dam structure as a part of the side walls, all incorporated as bed-slope source terms within the DG2 model discretisation. Simulations are run up to $t = 30$ s, during which the time histories of the water level and the velocity are computed at sampling points G2, G4 and G5 located around the building, where highly energetic flow features occur, and at point G6 behind the dam wall, where water flow is relatively slower. The time histories of the water level produced by DG2-LL and DG2-NL and the industrial models at resolution 0.1 m, and by DG2-LL and MUSCL-FV2 at resolution 0.2 m, are compared in Fig. 18 against the measured data from Soares-Frazão and Zech (2007).

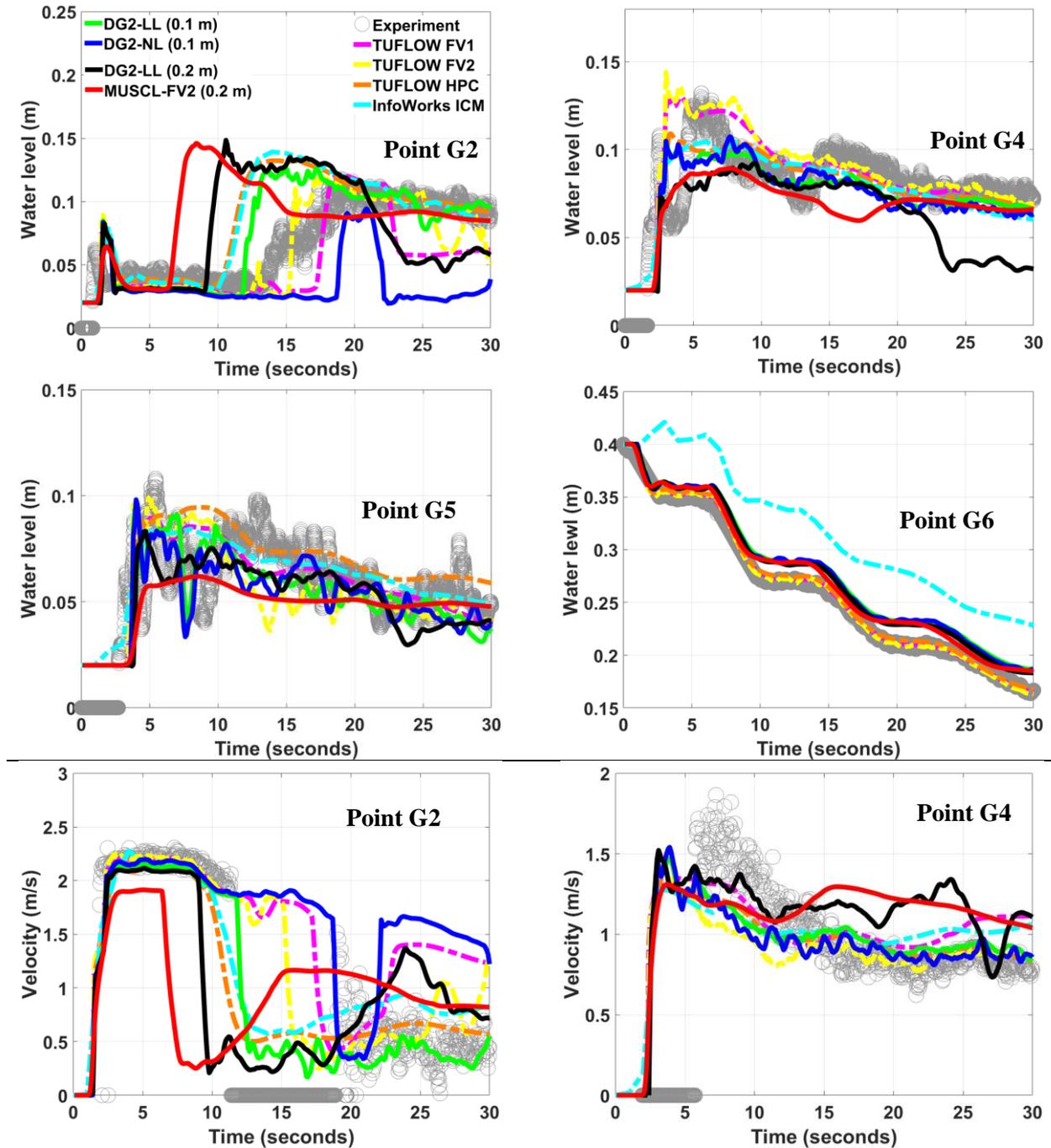

**Fig. 18.** Dam break over an oblique building. Time histories of the water level (top and middle parts) at points G2, G4, G5 and G6, and the velocity (bottom part) at points G2 and G4 relative to DG2-LL and DG2-NL at 0.1 m resolution, DG2-LL and MUSCL-FV2 models at 0.2 m resolution, the industrial models at 0.1 m resolution and the experimental data.

The water level and velocity histories at points G2 and G4 reveal that none of the models can fully reproduce the experimental profiles. At point G2, only the first wave arrival is well captured (around 1 sec) by the models, as it is clearer in the velocity predictions (Fig. 18, bottom panel, left part). Within the first 10-20 sec, the models fail to capture the rapidly increasing water jump transition (Fig. 18, top panel, left part), although the velocity histories (Fig. 18, bottom panel, left part) indicates that DG2-LL at 0.1 m resolution is the best performer, particularly in capturing the vanishing velocity at $t \sim 11$ s. In contrast, DG2-NL at a 0.1 m resolution fails to predict meaningful results in both water level and velocity histories (Fig. 18, top and bottom panels, left part), which is rather an expected shortcoming when swiching-off the slope limiter within a shock-dominated flow simulations. At point G4, the predicted depth and velocity histories indicate more clearly that none of the models can appropriately reproduce the experimental wave arrival times, which are smeared out (Fig. 18, top panel, right part) and predicted with a significant lag (Fig. 18, bottom panel, right part). These shortcomings in model predictions are not surprising as the wave patterns around points G2 and G4 become complex and highly unstable to be captured by 2D averaged SWE-based models under the hydrostatic assumption and without a turbulence closure. Nonetheless, the time histories at G2 and G4 still show benefits in applying DG2-LL to sufficiently reproduce small-scale transient features occurring in a highly dynamic flow type.

The DG2-LL capability is clearly discernible in the water level histories recorded at point G5 (Fig. 18, middle panel, left part). There, DG2-LL at a 0.1 m resolution excels in capturing the varying disturbances observed in the experimental histories with greater accuracy as compared to the industrial models, even at 0.2 m resolution where MUSCL-FV2 yields a relatively flat histories without disturbances. At point G6, where the flow is very much flood-like, the DG2 variants at 0.2 m resolution perform equally well as other models at 0.1 m resolution, thus demonstrating that the DG2 model can be a viable option to predict flows away from the shock-dominated area at twice-coarser resolution than those of the FV-based models. The better fit of the TUFLOW model predictions with the measured data can be attributed to the fact that these models have undergone ad-hoc calibration against the experimental data (Syme, 2008).

Fig. 19 contains the plots of the $R^2$ coefficients and $L^1$-norm errors at the respective sampling points. Except for point G6, the $R^2$ coefficients for the water level are at best around 0.6. This suggests that all models reproduce history patterns that are more than 60% distorted from the experimental patterns, although without a significant deviation since the $L^1$-norm errors do not exceed 0.06. On the other hand, the $R^2$ coefficients for the velocities are more illustrative of the DG2-LL's potential in capturing the transient patterns in a fast and abrupt flow ($R^2 > 0.6$), leading to the least deviated results (smallest $L^1$-norm error). Overall, this test recommends DG2-LL model to simulate complex flow hydrodynamics with shocks and to be particularly useful in the capturing of small-scale transients around eddy wakes.

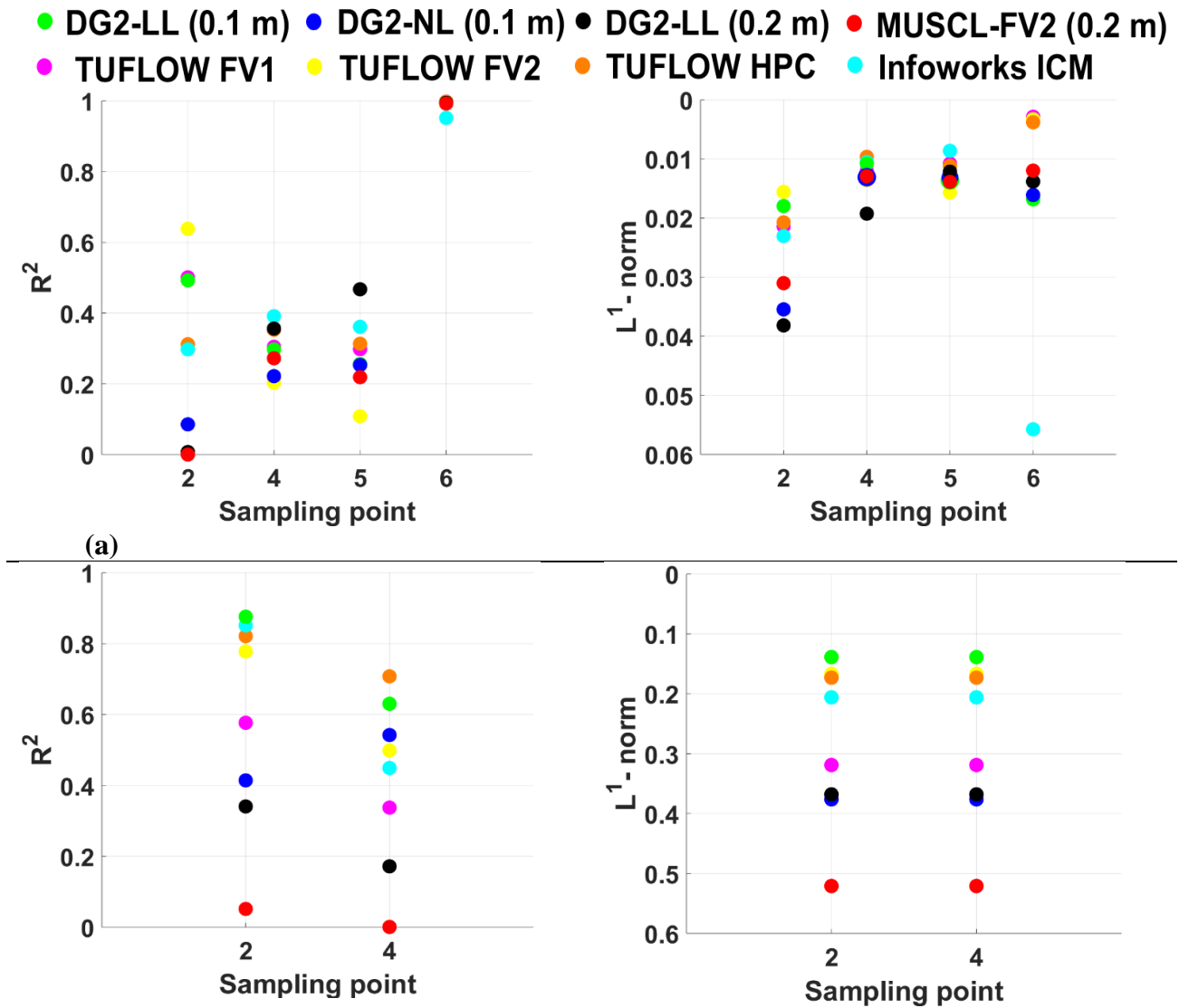

**(b)**

**Fig. 19.** Dam break over an oblique building. Difference among the flood model predictions relative to the experimental data: $R^2$ coefficients (left) and $L^1$-norm errors (right); (a) water level and (b) velocity.

### 3.3. Computational runtimes and discussions

For each test case in Secs 3.1 and 3.2, the runtimes generated by DG2-NL, DG2-LL and MUSCL-FV2 are recorded at output time $T$ as reported for the industrial models (Table 1) and for the best performer DG2 variant (DG2-BP) on a twice-coarser resolution. These runtimes, listed in Table 3, are recorded on the same desktop computer using a single-core CPU. Note that these runtimes should not be compared with those reported for the industrial models, which were either obtained on a CPU with 12 cores or on a GPU card with thousands of CUDA cores (hence excluded from this comparison).

**Table 3.** Runtime costs for DG2-NL, DG2-LL and MUSCL-FV2 based on the same grid resolution reported for the industrial models (Table 1), and considering the best performer DG2 variant (DG2-BP) on twice-coarser resolution. All simulation runtimes are generated from the same desktop computer using a single-core CPU.

| Test cases | Output time $T$ (sec) | Runtimes (sec): same resolution reported for the industrial models | | | Runtimes (sec): twice-coarser resolution |
|---|---|---|---|---|---|
| | | DG2-LL | DG2-NL | MUSCL-FV2 | DG2-BP |
| Flooding and drying cycle over a sloping topography (Sec. 3.1.1) | 72,000 (20 hrs) | 166 | 75 | 69 | 21 |
| Slow filling of multiple ponds (Sec. 3.1.2) | 172,000 (48 hrs) | 1,734 | 833 | 307 | 222 |
| Momentum conservation over an obstruction (Sec. 3.2.1) | 900 (15 min) | 12 | 5.5 | 3.4 | 1.5 |
| Torrential flooding over a rural catchment (Sec. 3.2.2) | 108,000 (30 hrs) | 22,260 | 12,297 | 3,822 | 2,320 |
| Dam break over an oblique building (Sec. 3.2.3) | 120 (2 min) | 15,337 | 10,363 | 1,205 | 497 |

As shown in Table 3, the runtimes of DG2-NL are found to be 1.8 to 2.2 times faster than those of DG2-LL in the first four test cases involving flood-like flows (Sec. 3.1.1 to Sec. 3.2.2). Compared to the runtimes of MUSCL-FV2 on the same resolution as the industrial models, DG2-NL is found to be 1.4 to 6.6 faster on twice-coarser resolution. Hence, DG2-NL is undoubtedly the best performer to support realistic flood modelling over large spatial areas and long timescales. This is also supported by the findings throughout Sections 3.1 and 3.2. DG2-NL partly falls short around the zone of shock-dominance (Sec. 3.2.3) and has the least speed up compared to DG2-LL (i.e. 1.4 times). For the simulation of flooding scenarios with shocks and wave reflection, e.g. a tsunami or dam-break wave propagation over across obstacles, DG2-LL should still be favoured as the best performer.

## 4. Summary and conclusions

This paper studied the potential of a second-order discontinuous Galerkin (DG2) model for practical flood modelling compared to industry-standard flood models, which are based on second-order finite volume (FV2) or first-order finite volume (FV1) numerical schemes. A simplified form of the DG2 flood model that is devised for robust flood modelling was presented, while drawing on its key (methodological) similarities and differences to the standard MUSCL-FV2 approach. Results from a synthetic test revealed notable momentum conservation property of the DG2 model due to its greater ability to withstand numerical dissipation, despite the grid resolution coarsening and long duration of simulation. This analysis also shows that the DG2 model with no slope limiting (DG2-NL) is as valid as the expensive variant with local slope limiting (DG2-LL) in simulating flood-like flows with wetting and drying movement that are often present in flood inundation modelling.

The performance of the DG2-LL and DG2-NL models was then explored for five test cases published by the UK Environment Agency, which are well-accepted by both industrial and academic hydraulic modelling communities for benchmarking the capabilities of 2D flood inundation models. The outputs of DG2-LL and DG2-NL were analysed and compared to the outputs of four prominent FV-based industrial models (i.e. TUFLOW-FV1, TUFLOW-FV2, TUFLOW-HPC and Infoworks ICM) for the same resolution reported for the industrial models. Outputs of the best performing DG2 variant (DG2-BP) were then analysed at twice-coarser resolution compared to the outputs of an in-house MUSCL-FV2 model. All analyses are quantified using $R^2$ coefficient and $L^1$-norm errors, representing the correlation and deviation of the model outputs with respect to a reference solution, either produced by running MUSCL-FV2 on the finest DEM available or via available measurement data. CPU runtimes were also scrutinised to contrast the runtime cost of DG2-LL vs. DG2-NL on the same resolution as the industrial models, and also DG2-BP vs. MUSCL-FV2 on a twice-coarser resolution.

The performance comparison between DG2 and FV-based model outputs offered a deeper insight into the practical requirements and the potential for deploying a grid-based DG2 flood model for industry-scale simulations. Local limiting was found to be unnecessary for a DG2 model to reliably support the modelling of a wide range of flooding flows (i.e. not dominated by strong flow discontinuities) such as in pluvial, fluvial and coastal flooding. The DG2-LL costs twice as much to run (Sec. 3.3) and can still produce misleading results, for example when a gradually-propagating inflow is still loading onto a small wet domain (Sec. 3.1.1). The DG2-NL yields closer predictions to the industrial model outputs, even at twice-coarser resolution. The DG2-NL at twice-coarser resolution could maintain a good correlation to the industrial models in simulating a flood flow driven by topographic- and friction-effects over a larger area (Sec. 3.1.2) and sample more accurately the water level hydrographs far away from the flood source.

For rapidly propagating flood waves on downslope channel with ponding (Sec. 3.2.1) and over a real-world natural valley (Sec. 3.2.2), DG2-NL, including at twice-coarser resolution, has shown ability to capture more detailed flow hydrographs incorporating small-scale transients. In particular, DG2-NL displayed sharper velocity peak and distinct oscillatory patterns of a sloshing flow settling in the ponds (Sec. 3.2.1) and managed to capture small transient curving in the velocity predictions (Sec. 3.2.2). In contrast, for a case with a strong dam-break wave interacting with a building-like obstacle (Sec. 3.2.3), deploying DG2-LL may still be useful for detailed capturing of small-scale flow variations caused by highly dynamic wave interactions with obstacles. Nonetheless, away from the shock-dominated region e.g. behind the dam wall (Sec. 3.2.3), DG2-NL can still produce competitive results relative to the industrial model outcomes. Overall, these findings lead to the conclusion that DG2-NL can offer a more efficient and accurate alternative for the industrial modelling communities to improve long-range and coarse-resolution flood flow simulations.

**References**

Alcrudo, F., 2004. A State of the Art Review on Mathematical Modelling of Flood Propagation. IMPACT Proj. 1–22.

Alkema, D., 2007. Simulating floods - on the application of a 2D-hydraulic model for flood hazard and risk assessment. University of Utrecht.

An, H., Yu, S., 2014. An accurate multidimensional limiter on quadtree grids for shallow water flow simulation. J. Hydraul. Res. 52, 565–574. https://doi.org/10.1080/00221686.2013.878404

Arrighi, C., Pregnolato, M., Dawson, R.J., Castelli, F., 2019. Science of the Total Environment Preparedness against mobility disruption by fl oods. Sci. Total Environ. 654, 1010–1022. https://doi.org/10.1016/j.scitotenv.2018.11.191

Ayog, J.L., Kesserwani, G., 2020. Modelling results for the analytical assessment (Sec. 2.3) and EA benchmark tests (Sec. 3) [Datasets]. https://doi.org/10.5281/zenodo.3760628

Bai, F.P., Yang, Z.H., Zhou, W.G., 2018. Study of total variation diminishing (TVD) slope limiters in dam-break flow simulation. Water Sci. Eng. 11, 68–74. https://doi.org/10.1016/j.wse.2017.09.004

Berger, M., Aftosmis, M.J., Murman, S.M., 2005. Analysis of Slope Limiters on Irregular Grids, in: 43rd AIAA Aerospace Sciences Meeting. Reno, Nevada, pp. 1–22.

Bladé, E., Cea, L., Corestein, G., Escolano, E., Puertas, J., Vázquez-Cendón, E., Dolz, J., Coll, A., 2014. Iber: herramienta de simulación numérica del flujo en ríos. Rev. Int. Métodos Numéricos para Cálculo y Diseño en Ing. 30, 1–10. https://doi.org/https://doi.org/10.1016/j.rimni.2012.07.004

BMT-WBM, 2018. TUFLOW Classic/HPC User Manual Build 2018-03-AD.

BMT-WBM, 2016. TUFLOW User Manual Build 2016-03-AE.

Bunya, S., Kubatko, E.J., Westerink, J.J., Dawson, C., 2009. A wetting and drying treatment for the Runge–Kutta discontinuous Galerkin solution to the shallow water equations. Comput. Methods Appl. Mech. Eng. 198, 1548–1562. https://doi.org/10.1016/j.cma.2009.01.008

Cheng, C., Qian, X., Zhang, Y., 2011. Estimation of the evacuation clearance time based on dam-break simulation of the Huaxi dam in Southwestern China. Nat. Hazards 57, 227–243. https://doi.org/10.1007/s11069-010-9608-4

Clare, M., Percival, J., Angeloudis, A., Cotter, C., Piggott, M., 2020. Hydro-morphodynamics 2D modelling using a discontinuous Galerkin discretisation. https://doi.org/10.31223/OSF.IO/TPQVY

Cockburn, B., Shu, C., 2001. Runge-Kutta Discontinuous Galerkin Methods for Convection Dominated Problems. J. Sci. Comput. 16, 173–261.

Costabile, P., Costanzo, C., De Lorenzo, G., Macchione, F., 2020. Is local flood hazard assessment in urban areas significantly influenced by the physical complexity of the hydrodynamic inundation model? J. Hydrol. 580. https://doi.org/10.1016/j.jhydrol.2019.124231

Crossley, A. Lamb, R. Waller, S., 2010. Fast solution of the shallow water equations using GPU technology, in: BHS Third International Symposium: Managing Consequences of a Changing Global Environment. Newcastle, United Kingdom. https://doi.org/10.1017/CBO9781107415324.004

de Almeida, G.A.M., Bates, P., Ozdemir, H., 2018. Modelling urban floods at submetre resolution: challenges or opportunities for flood risk management? J. Flood Risk Manag. 11, S855–S865. https://doi.org/10.1111/jfr3.12276

Delis, A.I., Nikolos, I.K., 2013. A novel multidimensional solution reconstruction and edge-based limiting procedure for unstructured cell-centered finite volumes with application to shallow water dynamics. Int. J. Numer. Methods Fluids 71, 584–633. https://doi.org/10.1002/fld.3674

Delis, A.I., Nikolos, I.K., Kazolea, M., 2011. Performance and Comparison of Cell-Centered and Node-Centered Unstructured Finite Volume Discretizations for Shallow Water Free Surface Flows. Arch Comput Methods Eng. 18, 57–118. https://doi.org/10.1007/s11831-011-9057-6

Echeverribar, I., Morales-Hernández, M., Brufau, P., García-Navarro, P., 2019. 2D numerical simulation of unsteady flows for large scale floods prediction in real time. Adv. Water Resour.

134, 103444. https://doi.org/10.1016/j.advwatres.2019.103444

Engineers Australia, 2012. Australian Rainfall & Runoff Revision Project 15: Two Dimensional Modelling in Urban and Rural floodplains, Stage 1 and 2 Draft Report, Engineers Australia.

Ern, A., Piperno, S., Djadel, K., 2008. A well-balanced Runge-Kutta discontinuous Galerkin method for the shallow-water equations with flooding and drying. Int. J. Numer. Methods Fluids 1–25. https://doi.org/10.1002/fld

Fu, G., Shu, C.-W., 2017. A new troubled-cell indicator for discontinuous Galerkin methods for hyperbolic conservation laws. J. Comput. Phys. 347, 305–327. https://doi.org/https://doi.org/10.1016/j.jcp.2017.06.046

Ginting, B.M., 2019. Central-upwind scheme for 2D turbulent shallow flows using high-resolution meshes with scalable wall functions. Comput. Fluids 179, 394–421. https://doi.org/10.1016/j.compfluid.2018.11.014

Guard, P., Nielsen, C., Ryan, P., Teakle, I., 2013. Parameter sensitivity of a 2D finite volume hydrodynamic model and its application to tsunami simulation.

Henckens, G., Engel, W., 2017. Benchmark inundatiemodellen: modelfunctionaliteiten en testbank berekeningen. Amersfoort, The Netherlands.

Horváth, Z., Buttinger-Kreuzhuber, A., Konev, A., Cornel, D., Komma, J., Blöschl, G., Noelle, S., Waser, J., 2020. Comparison of Fast Shallow-Water Schemes on Real-World Floods. J. Hydraul. Eng. 146, 1–16. https://doi.org/10.1061/(ASCE)HY.1943-7900.0001657

Hou, J., Liang, Q., Zhang, H., Hinkelmann, R., 2015. An efficient unstructured MUSCL scheme for solving the 2D shallow water equations. Environ. Model. Softw. 66, 131–152. https://doi.org/10.1016/j.envsoft.2014.12.007

Huxley, C., Syme, B., Symons, E., 2017. UK Environment Agency 2D Hydraulic Model Benchmark Tests. 2017-09 TUFLOW Release Update. Brisbane, Australia.

Hydronia LLC, 2019. HydroBID Flood Two-Dimensional Flood and River Dynamics Model.

Innovyze, 2011. Infoworks 11.5 RS Help. Broomfield, Colorado.

Jamieson, S.R., Lhomme, J., Wright, G., Gouldby, B., 2012. A highly efficient 2D flood model with sub-element topography. Proc. Inst. Civ. Eng. Water Manag. 165, 581–595. https://doi.org/http://dx.doi.org/10.1680/wama.12.00021

Kärnä, T., Kramer, S.C., Mitchell, L., Ham, D.A., Piggott, M.D., Baptista, A.M., 2018. Thetis coastal ocean model: discontinuous Galerkin discretization for the three-dimensional hydrostatic equations. Geosci. Model Dev. Discuss. 1–36. https://doi.org/10.5194/gmd-2017-292

Kesserwani, G., Ayog, J.L., Bau, D., 2018. Discontinuous Galerkin formulation for 2D hydrodynamic modelling: Trade-offs between theoretical complexity and practical convenience. Comput. Methods Appl. Mech. Eng. 342, 710–741. https://doi.org/10.1016/j.cma.2018.08.003

Kesserwani, G., Ghostine, R., Vazquez, J., Ghenaim, A., Mosé, R., 2008. Riemann Solvers with Runge–Kutta Discontinuous Galerkin Schemes for the 1D Shallow Water Equations. J. Hydraul. Eng. 134, 243–255. https://doi.org/10.1061/(ASCE)0733-9429(2008)134:2(243)

Kesserwani, G., Liang, Q., 2012a. Influence of Total-Variation-Diminishing Slope Limiting on Local Discontinuous Galerkin Solutions of the Shallow Water Equations. J. Hydraul. Eng. 138, 216–222. https://doi.org/10.1061/(ASCE)HY.1943-7900.0000494

Kesserwani, G., Liang, Q., 2012b. Locally limited and fully conserved RKDG2 shallow water solutions with wetting and drying. J. Sci. Comput. 50, 120–144. https://doi.org/10.1007/s10915-011-9476-4

Kesserwani, G., Liang, Q., 2011. A conservative high-order discontinuous Galerkin method for the shallowwater equations with arbitrary topography. Int. J. Numer. Methods Eng. 86, 47–69. https://doi.org/10.1002/nme

Kesserwani, G., Liang, Q., 2010. Well-balanced RKDG2 solutions to the shallow water equations over irregular domains with wetting and drying. Comput. Fluids 39, 2040–2050. https://doi.org/10.1016/j.compfluid.2010.07.008

Kesserwani, G., Sharifian, M.K., 2020. (Multi)wavelets increase both accuracy and efficiency of standard Godunov-type hydrodynamic models: Robust 2D approaches. Adv. Water Resour. 144. https://doi.org/10.1016/j.advwatres.2020.103693

Kesserwani, G., Wang, Y., 2014. Discontinuous galerkin flood model formulation: Luxury or necessity? Water Resour. Res. 50, 6522–6541. https://doi.org/10.1002/2013WR014906

Krivodonova, L., Xin, J., Remacle, J.F., Chevaugeon, N., Flaherty, J.E., 2004. Shock detection and limiting with discontinuous Galerkin methods for hyperbolic conservation laws. Appl. Numer. Math. 48, 323–338. https://doi.org/10.1016/j.apnum.2003.11.002

Kubatko, E.J., Westerink, J.J., Dawson, C., 2006. hp Discontinuous Galerkin methods for advection dominated problems in shallow water flow. Comput. Methods Appl. Mech. Eng. 196, 437–451. https://doi.org/10.1016/j.cma.2006.05.002

Lambrechts, J., Humphrey, C., McKinna, L., Gourge, O., Fabricius, K.E., Mehta, A.J., Lewis, S., Wolanski, E., 2010. Importance of wave-induced bed liquefaction in the fine sediment budget of Cleveland Bay, Great Barrier Reef. Estuar. Coast. Shelf Sci. 89, 154–162. https://doi.org/10.1016/j.ecss.2010.06.009

Latrubesse, E.M., Park, E., Sieh, K., Dang, T., Lin, Y.N., Yun, S., 2020. Geomorphology Dam failure and a catastrophic fl ood in the Mekong basin ( Bolaven. Geomorphology 362, 107221. https://doi.org/10.1016/j.geomorph.2020.107221

Le Bars, Y., Vallaeys, V., Deleersnijder, É., Hanert, E., Carrere, L., Channelière, C., 2016. Unstructured-mesh modeling of the Congo river-to-sea continuum. Ocean Dyn. 66, 589–603. https://doi.org/10.1007/s10236-016-0939-x

Le, H.A., Lambrechts, J., Ortleb, S., Gratiot, N., Deleersnijder, E., Soares-Frazão, S., 2020. An implicit wetting–drying algorithm for the discontinuous Galerkin method: application to the Tonle Sap, Mekong River Basin. Environ. Fluid Mech. https://doi.org/10.1007/s10652-019-09732-7

Lhomme, J., Gutierrez-Andres, J., Weisgerber, A., Davison, M., Mulet-Marti, J., Cooper, A.,

Gouldby, B., 2010. Testing a new two-dimensional flood modelling system: analytical tests and application to a flood event. J. Flood Risk Manag. 3, 33–51. https://doi.org/10.1111/j.1753-318X.2009.01053.x

Liang, Q., Marche, F., 2009. Numerical resolution of well-balanced shallow water equations with complex source terms. Adv. Water Resour. 32, 873–884. https://doi.org/10.1016/j.advwatres.2009.02.010

Marras, S., Kopera, M.A., Constantinescu, E.M., Suckale, J., Giraldo, F.X., 2018. A residual-based shock capturing scheme for the continuous/discontinuous spectral element solution of the 2D shallow water equations. Adv. Water Resour. 114, 45–63. https://doi.org/10.1016/j.advwatres.2018.02.003

McRae, A.T.T., 2015. Compatible finite element methods for atmospheric dynamical cores (PhD thesis). Imperial College London.

Mulamba, T., Bacopoulos, P., Kubatko, E.J., Pinto, G.F., 2019. Sea-level rise impacts on longitudinal salinity for a low-gradient estuarine system. Clim. Change 152, 533–550. https://doi.org/10.1007/s10584-019-02369-x

Mungkasi, S., Roberts, S.G., 2013. Validation of ANUGA hydraulic model using exact solutions to shallow water wave problems. J. Phys. Conf. Ser. 423. https://doi.org/10.1088/1742-6596/423/1/012029

Neal, J., Villanueva, I., Wright, N., Willis, T., Fewtrell, T., Bates, P., 2012. How much physical complexity is needed to model flood inundation? Hydrol. Process. 26, 2264–2282. https://doi.org/10.1002/hyp.8339

Neelz, S., Pender, G., 2010. Benchmarking of 2D Dyfraulic Modelling Packages. DEFRA/Environment Agency, UK.

Neelz, S., Pender, G., 2013. Benchmarking the Latest Generation of 2D Hydraulic Modelling Packages. DEFRA/Environment Agency, UK.

Neelz, S., Pender, G., 2009. Desktop Review of 2D Hydraulic Modelling Packages.

DEFRA/Environment Agency, UK. DEFRA/Environment Agency, UK.

Patel, D.P., Ramirez, J.A., Srivastava, P.K., 2017. Assessment of flood inundation mapping of Surat city by coupled 1D / 2D hydrodynamic modeling : a case. Nat. Hazards 89, 93–130. https://doi.org/10.1007/s11069-017-2956-6

Pham Van, C., de Brye, B., Deleersnijder, E., Hoitink, A.J.F., Sassi, M., Spinewine, B., Hidayat, H., Soares-Frazão, S., 2016. Simulations of the flow in the Mahakam river–lake–delta system, Indonesia. Environ. Fluid Mech. 16, 603–633. https://doi.org/10.1007/s10652-016-9445-4

Pistrika, A.K., Jonkman, S.N., 2010. Damage to residential buildings due to flooding of New Orleans after hurricane Katrina. Nat. Hazards 54, 413–434. https://doi.org/10.1007/s11069-009-9476-y

Qiu, J., Shu, C., 2005. A Comparison of Troubled-Cell Indicators for Runge-Kutta Discontinuous Galerkin Methods Using Weighted Essentially Nonoscillatory Limiters. SIAM J. Sci. Comput. 27, 995–1013.

Reis, C., Figueiredo, J., Clain, S., Omira, R., Baptista, M.A., Miranda, J.M., 2019. Comparison between MUSCL and MOOD techniques in a finite volume well-balanced code to solve SWE. the Tohoku-Oki, 2011 example. Geophys. J. Int. 216, 958–983. https://doi.org/10.1093/gji/ggy472

Sanders, B.F., Bradford, S.F., 2006. Impact of limiters on accuracy of high-resolution flow and transport models. J. Eng. Mech. 132, 87–98. https://doi.org/10.1061/(ASCE)0733-9399(2006)132:1(87)

Schubert, J.E., Sanders, B.F., Smith, M.J., Wright, N.G., 2008. Unstructured mesh generation and landcover-based resistance for hydrodynamic modeling of urban flooding. Adv. Water Resour. 31, 1603–1621. https://doi.org/10.1016/j.advwatres.2008.07.012

Shaw, J., Kesserwani, G., Neal, J., Bates, P., Sharifian, M.K., 2020. LISFLOOD-FP 8 . 0 : the new discontinuous Galerkin shallow water solver for multi-core CPUs and GPUs. Geosci. Model Dev. Discuss. (in review)

Shirvani, M., Kesserwani, G., Richmond, P., 2020a. Agent-based modelling of pedestrian responses during flood emergency: mobility behavioural rules and implications for flood risk analysis. J. Hydroinformatics 22, 1078–1092. https://doi.org/https://doi.org/10.2166/hydro.2020.031

Shirvani, M., Kesserwani, G., Richmond, P., 2020b. Agent-based simulator of dynamic flood-people interactions. http://arxiv.org/abs/1908.05232.

Soares-Frazão, S., Zech, Y., 2008. Dam-break flow through an idealised city. J. Hydraul. Res. 46, 648–658. https://doi.org/10.3826/jhr.2008.3164

Soares-Frazão, S., Zech, Y., 2007. Experimental study of dam-break flow against an isolated obstacle. J. Hydraul. Res. 45, 27–36. https://doi.org/10.1080/00221686.2007.9521830

Sweby, P.K., 1984. High Resolution Schemes Using Flux Limiters for Hyperbolic Conservation Laws. SIAM J. Numer. Anal. 21, 995–1011.

Syme, W.J., 2008. Flooding in Urban Areas - 2D Modelling Approaches for Buildings and Fences, in: 9th National Conference on Hydraulics in Water Engineering: Hydraulics. Engineers Australia, Barton, A.C.T., pp. 25–32.

Teng, J., Jakeman, A.J., Vaze, J., Croke, B.F.W., Dutta, D., Kim, S., 2017. Flood inundation modelling: A review of methods, recent advances and uncertainty analysis. Environ. Model. Softw. 90, 201–216. https://doi.org/10.1016/j.envsoft.2017.01.006

Vater, S., Beisiegel, N., Behrens, J., 2019. A limiter-based well-balanced discontinuous Galerkin method for shallow-water flows with wetting and drying: Triangular grids. Int. J. Numer. Methods Fluids 91, 395–418. https://doi.org/10.1002/fld.4762

Vetsch, D., Siviglia, A., Caponi, F., Ehrbar, D., Gerke, E., Kammerer, S., Koch, A., Peter, S., Vanzo, D., Vonwiller, L., Facchini, M., Gerber, M., Volz, C., Farshi, D., Mueller, R., Rousselot, P., Veprek, R., Faeh, R., 2018. System Manuals of BASEMENT, Version 2.8. Laboratory of Hydraulics, Glaciology and Hydrology (VAW).

Wood, D., Kubatko, E.J., Rahimi, M., Shafieezadeh, A., Conroy, C.J., 2020. Implementation and evaluation of coupled discontinuous Galerkin methods for simulating overtopping of flood defenses by storm waves. Adv. Water Resour. 136. https://doi.org/10.1016/j.advwatres.2019.103501

Xia, X., Liang, Q., Ming, X., 2019. A full-scale fluvial flood modelling framework based on a high-performance integrated hydrodynamic modelling system (HiPIMS). Adv. Water Resour. 132, 103392. https://doi.org/10.1016/j.advwatres.2019.103392

Yoon, T.H., Kang, S.-K., 2004. Finite Volume Model for Two-Dimensional Shallow Water Flows on Unstructured Grids. J. Hydraul. Eng. 130, 678–688. https://doi.org/10.1061/(ASCE)0733-9429(2004)130

Zhao, J., Özgen, I., Liang, D., Hinkelmann, R., 2018. Improved multislope MUSCL reconstruction on unstructured grids for shallow water equations. Int. J. Numer. Methods Fluids 87, 401–436. https://doi.org/10.1002/fld.4499

**Acknowledgements and DG2 software accessibility**

Georges Kesserwani, James Shaw and Mohammad Kazem Sharifian acknowledge the support of the UK Engineering and Physical Sciences Research Council, grant ID: EP/R007349/1; and, Janice Lynn Ayog acknowledges the support of the Ministry of Higher Education, Malaysia and Universiti Malaysia Sabah, Malaysia. We wish to thank Julien Lhomme (Innovyze) and Duncan Kitts (BMT-WBM) for providing the outputs of Infoworks ICM and TUFLOW models for the UK Environment Agency benchmark test cases. We also thank Paul Bates (University of Bristol), Bill Syme (BMT-WBM), for their feedbacks and insightful comments on the comparative analyses in Sec. 3. The DG2-flood model has been integrated into LISFLOOD-FP where it can be freely accessed by following the instructions reported in Shaw et al. (2020).

**CRediT authorship contribution statement**

**Janice Lynn Ayog**: Methodology, Software, Visualization, Validation, Investigation, writing - original draft, review and editing. **Georges Kesserwani**: Conceptualisation, Supervision, Formal analysis, Writing - original draft, review and editing. **James Shaw**: Conceptualisation, Software, Writing - review and editing. **Mohammad Kazem Sharifian**: Conceptualisation, Writing - review and editing. **Domenico Bau**: Supervision, Writing - review and editing.